\documentclass[
 reprint,
 preprintnumbers,
 amsmath,amssymb,
 aps, 
 pre, 
 superscriptaddress,
 floatfix
]{revtex4-2}

\usepackage{graphicx}
\usepackage{dcolumn}
\usepackage{bm}
\usepackage{xcolor}
\usepackage[normalem]{ulem}
\begin{document}

\preprint{Accepted for publication in Physical Review E}

\title{Prospective Coding and Path Integration Emerge as Equilibrium Solutions of Self-Organizing Neural Networks with Firing-Rate Adaptation}

\author{Facundo Emina}
 \email{femina@leloir.org.ar}
\affiliation{Universidad de Buenos Aires, Facultad de Ciencias Exactas y Naturales, Departamento de Física. Buenos Aires, Argentina.}
\affiliation{Leloir Institute---IIBBA/CONICET. Buenos Aires, Argentina.}

\author{Emilio Kropff}
 \email{ekropff@leloir.org.ar}
\affiliation{Leloir Institute---IIBBA/CONICET. Buenos Aires, Argentina.}

\date{\today}

\begin{abstract}
Continuous Attractor Neural Networks (CANNs) traditionally rely on pre-wired recurrent connectivity to model spatial representations, path integration, and anticipatory dynamics. However, the biological mechanisms through which this structured connectivity emerges via learning remain relatively unexplored. This work presents a theoretical framework revealing how continuous attractor connectivity and its computational properties self-organize through Hebbian plasticity, firing-rate adaptation, and global inhibition. We show that translationally invariant inputs naturally drive the emergence of stable, Gaussian-profiled feedforward weights. Crucially, anticipatory dynamics arise spontaneously within these feedforward architectures, shifting the activity bump forward without requiring recurrent excitatory collaterals. This predictive shift can be linearly amplified across multilayer networks, consistent with anticipatory activity observed in the superficial layers of the entorhinal cortex. Furthermore, introducing recurrent interactions allows the network to learn connections capable of self-sustaining a moving bump of activity. Finally, by modulating the network with an external, time-varying baseline current that encodes speed, the system adjusts its intrinsic velocity to function as a precise unidirectional path integrator. Ultimately, this study suggests that prospective coding and path integration are not manually engineered features, but rather naturally co-emergent properties of a single self-organizing competitive network.
\end{abstract}

\maketitle

\section{Introduction}

Attractor neural networks have long served as canonical models for understanding how neural populations store and process information. In the classical formulation, Hopfield networks encode discrete patterns as stable fixed-point attractors arising from Hebbian synaptic plasticity \citep{hopfield1982neural}. Continuous attractor neural networks (CANNs) extend this framework to continuous variables—such as spatial position or orientation—by sustaining localized activity bumps that evolve along low-dimensional manifolds. This formulation has proven effective in modeling a variety of neural systems, including multiple environment-specific representations of space in the hippocampus of the rodent brain \citep{battaglia1998attractor}, as well as grid-cell \citep{burak2009accurate}, and head-direction responses \citep{skaggs1994model,redish1996coupled}. In these architectures, the population activity manifests as a localized bump on the representational manifold, where individual neurons fire only when the encoded variable falls within their specific receptive fields.

Many of these systems support computations that require the internal representation to evolve coherently in time, such as prospective coding and path integration—the ability to estimate position or orientation by integrating self-motion cues. Maintaining a representation that leads the current input is essential for both real-time control and long-term planning. While this mechanism is often understood as a way to compensate for processing delays \citep{nijhawan2009compensating}, the utility of anticipation extends further; by representing future states, the system can proactively prepare for incoming stimuli and facilitate planning.

Consistent with this requirement, anticipatory activity has been extensively documented across a wide variety of neural systems. For example, the visual system employs predictive mechanisms to compensate for delays during stimulus tracking \citep{duhamel1992updating,berry1999anticipation}. Similar anticipatory control is observed in birds during flight prior to water entry \citep{lee1981plummeting}, and in head-direction cells, which encode the future orientation of rodents with a lead time of approximately 25 ms \citep{blair1995anticipatory}.

In the head-direction system, anticipatory dynamics are closely linked to path integration \citep{sharp2001angular}. This computation is a hallmark of grid-cell and head-direction systems in the entorhinal cortex \citep{mcnaughton2006path,gil2017impairedpathintegration}. The prevailing view is that grid cells perform path integration by combining speed signals, encoded in the firing-rates of speed cells \citep{kropff2015speed}, with directional inputs \citep{sargolini2006conjunctive,zutshi2018recurrent}, while incorporating environmental cues for error correction \citep{hoydal2019object,solstad2008representation}. 

Prospective coding has also been reported in neural representations of both speed and position in the entorhinal cortex \citep{kropff2015speed,ouchi2024predictive}, suggesting that these two computations may be intrinsically related. However, in CANN models, path integration \citep{burak2009accurate} and anticipatory coding via firing-rate adaptation \citep{mi2014spike,yu2019spiking,li2024dynamics} are typically reproduced separately and generally rely on pre-wired recurrent architectures. 

Within this modeling framework, canonical CANNs typically rely on symmetric recurrent connectivity to sustain a static bump of activity capable of tracking a moving input. However, several extensions have shown that richer dynamical regimes can emerge from modest modifications to this class of models. Mechanisms such as asymmetric connectivity \citep{spalla2021continuous}, explicit inhibitory populations \citep{zhang2012neural,yan2023bimodular}, or intrinsic neuronal adaptation \citep{mi2014spike} can generate moving bumps and anticipatory dynamics, providing candidate mechanisms for these computations.

In parallel, analytical work has focused on tractable formulations of CANNs. A prominent example is the Gaussian connectivity model with quadratic activation function and divisive normalization \citep{deneve1999reading,wu2002population,fung2008dynamics,fung2010moving}, where the network dynamics can be diagonalized in a Hermite-polynomial basis. This approach enables substantial dimensionality reduction and allows for closed-form characterization of bump shape, stability, and dynamical response. However, this analytical tractability comes at the cost of assuming a predefined connectivity structure.

In contrast, the question of how such structured connectivity emerges through learning remains relatively unexplored. Models of trace learning and temporal sequence learning have a long history \citep{foldiak1991learning, sharp1991computer}. In these models, Hebbian plasticity is combined with a temporally extended trace of neural activity, so that inputs occurring close together in time become associated, allowing stable representations and invariances to self-organize from sequences of inputs. Related Hebbian learning rules have since been applied to continuous attractor systems—including grid-cell models \citep{fuhs2006spin,kropff2008emergence,widloski2014model}, motor-sequence networks \citep{SOCANN-Stringer-Rolls}, and head-direction networks \citep{ringattractor-TizianoKempter}. However, these studies are generally not analytically tractable and tend to emphasize functional outcomes over the conditions required for the emergence and stability of the underlying connectivity. As a result, a theoretical understanding of how continuous attractors arise from learning, and how their computational properties are shaped by plasticity, is still lacking.

In this work, we develop a theoretical framework to characterize the emergence of continuous attractor connectivity and its link to network function. Using grid-cell-like dynamics as a reference, we investigate how learning gives rise to both prospective coding and path integration. Following a similar approach to \cite{kropff2008emergence}, we first analyze the emergence of structured feedforward connectivity and show how it supports low-dimensional representations and generates anticipatory dynamics. Prospective coding arises spontaneously within feedforward architectures, without requiring recurrent excitatory collaterals, through intrinsic firing-rate adaptation, in line with earlier modeling work on hippocampal circuits \citep{treves2004computational}. This is of potential interest for the superficial layers of medial entorhinal cortex, where anticipatory activity has been reported \citep{kropff2015speed,ouchi2024predictive} and recurrent excitation is thought to be limited \citep{dhillon2000laminar}. We then incorporate recurrent interactions to study the emergence and stability of attractor states across parameter regimes. Finally, we provide an analytical treatment of path integration, focusing on the network’s ability to calibrate its internal velocity and maintain accurate position estimates.

\section{Results}

\subsection{\label{subsec:TheModel}The Model with Feedforward Connections}

We consider a two-layer neural network. The first, input, layer consists of $N_\text{in}$ neurons that encode a moving stimulus propagating in a single direction with constant velocity, $v$, along a segment of length $L$. Each neuron is tuned to exhibit a Gaussian response centered at a given position, with width $\sigma_R$. For simplicity, we identify a neuron by the position $x$ where its peak response is achieved. We assume homogeneous density along the segment, given by $\rho_\text{in}=N_\text{in}/L$. The moving stimulus, or tutor, follows a periodic trajectory starting at $x = 0$ each time it reaches $x = L$. Throughout this work, we assume  $L\gg \sigma_R$ in both our theoretical analysis and simulations, ensuring that integrals across the finite domain can be treated as integrals over the entire real line. Assuming the density is high enough to approximate the population by a continuous distribution, the activity at time $t$ of a neuron centered at position $x$ is given by

\begin{equation}
    R(x,t) = A_R\,  \mathcal{N}(x; \, \, vt, \sigma_R), \label{eq:moving-input-eq}
\end{equation}

\noindent where $\mathcal{N}(x; \, \,\mu,\sigma) = \exp\left[-\tfrac{(x-\mu)^2}{2\sigma^2}\right]/\sqrt{2\pi\sigma^2}$ denotes a Gaussian with mean $\mu$ and standard deviation $\sigma$. With this convention, the total drive to the layer is set by \(A_R\), while the Gaussian profile distributes this activity across neurons. 

A second layer of $N_c$ neurons and density $\rho_c$ receives input from the tutor layer. Its dynamics are that of a CANN model with firing-rate adaptation, as described in \cite{mi2014spike}:

\begin{align}
    \tau \frac{\partial }{\partial t} U(x,t)&= -U(x,t) - V(x,t) + I(x,t), 
\label{eq:U-dynamics}\\
    \tau_v \frac{\partial }{\partial t}V(x,t) &= -V(x,t) + m\, U(x,t),
\label{eq:V-dynamics}
\end{align}

\noindent where $U(x,t)$ denotes the membrane potential of a neuron at position $x$ and time $t$, $V(x,t)$ the firing-rate adaptation variable, and $m$ the strength of the adaptation. Note that $x$ does not necessarily represent the physical position of the external stimulus, but rather the coordinates of the neuron within its internal neural manifold. The term $I(x,t)$ is the total input current to the neuron, which may include both feedforward input from other layers and recurrent contributions within the layer itself, and can vary with $x$ if neurons receive different inputs. In this first section, we focus on the case where $I(x,t)$ is the feedforward drive from the input layer:

\begin{equation}
    I(x,t)  = \rho_{\text{in}} \int_{x^{\prime}}dx^{\prime}\, J(x,x^{\prime}) R(x^{\prime},t)\, , 
    \label{eq:input-current}
\end{equation}

\noindent where the integration is done over the interval $[0,L)$ and $J(x,x^{\prime})$ denotes the feedforward synaptic weight from neuron $x^{\prime}$ in the input layer to neuron $x$ in the competitive layer.

The firing-rate $r(x, t)$ of the neural population is modeled as a nonlinear function of the membrane potential $U(x, t)$, defined by:

\begin{equation}
    \begin{aligned}
        r(x, t) =& \frac{[U(x, t)]_+^2}{B},\\
        \text{where} \quad B =& 1 + k \rho_c \int_{x^{\prime}} dx^{\prime} \, [U(x^{\prime}, t)]_+^2,
    \end{aligned}
\end{equation}

\noindent and $[\cdot]_+$ rectifies the content taking the maximum value between it and zero. The firing-rate is subject to global divisive normalization, governed by the inhibition parameter $k$. The integration in the normalization term expression $B$ is defined over the entire domain without explicit spatial limits. This generalized notation allows the model to account for a wide range of neural codes, encompassing both initially unstructured (disorganized) populations and those that have acquired a specific spatial organization.

To incorporate learning, we introduce Hebbian plasticity in feedforward weights between the two layers. Building on the principle that temporal input sequences can drive the self-organization of stable spatial representations \citep{foldiak1991learning}, the weight update takes the form:

\begin{equation}\begin{split}
\frac{\partial }{\partial t}J(x,x^{\prime}) = \eta_J r(x,t) \left[ R(x^{\prime},t) - \alpha_J  J(x,x^{\prime})^\beta \right]
\end{split}
\label{eq:hebbian-learning}\end{equation}

\noindent where $\eta_J$ is the learning rate. We set $1/\eta_J\gg \tau_v$, so that plasticity evolves on a slower timescale than the neural dynamics. In this expression, the first term implements Hebbian potentiation: synapses are strengthened in proportion to the co-activation between presynaptic drive $R(x,t)$ and postsynaptic activity $r(x^{\prime},t)$. The second term introduces a decay mechanism that mimics a local homeostatic process, with the parameter $\beta >0$ controlling the extent and $\alpha_J>0$ the strength of this decay.

Finally, to respect Dale’s law, we restrict $J(x,x^{\prime})$ to excitatory connections. In simulations, any update that results in a negative weight is clipped at zero.

\subsection{\label{subsec:FFWeights}Self Organized Feedforward Connectivity}
\subsubsection{Gaussian shape}

\begin{figure*}[t]
    \centering
    \includegraphics[width=1\linewidth]{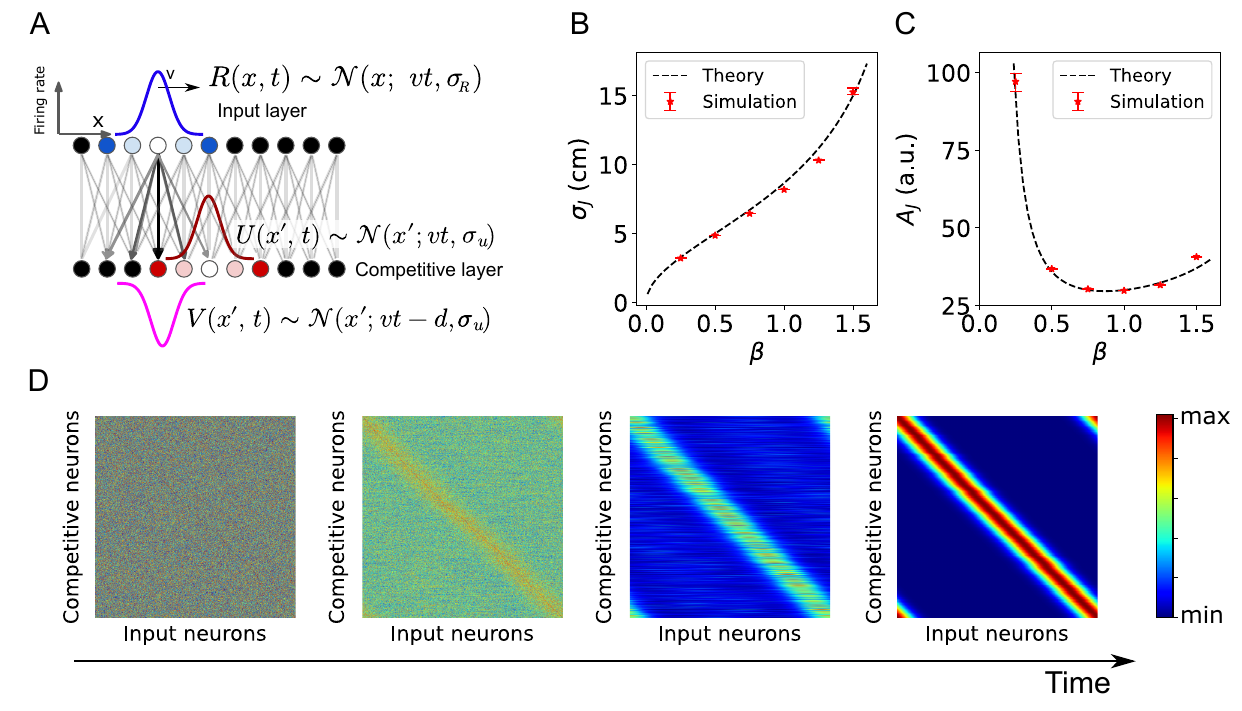}
    \caption{\textbf{Self-Organized Feedforward Connectivity}. \textbf A, Schematic of the two-layer model after learning. An input layer of neurons $R(x,t)$, encoding a unidirectionally moving stimulus  at speed $v$ through Gaussian tuning curves, projects through a Gaussian connectivity profile to a second layer $U(x,t)$ that implements competition via Hebbian learning, firing-rate adaptation, and global inhibition (see Eqs.~\ref{eq:U-dynamics},~\ref{eq:V-dynamics}~and~\ref{eq:hebbian-learning}). \textbf B, Learned standard deviation as a function of the $\beta$ parameter. Red: simulation median and interquartile ranges; black dashed line: theoretical prediction (Eq.~\ref{eq:consistent-sigmas}). Simulations were performed using the Euler method with a time step of $\Delta t=5\cdot 10^{-3}$ s. Parameters used: $N_\text{in}= N_c=512$, $k=5.12\cdot 10^{-2}$ $\text{cm}^4\text{s}^2/\text{syn}$, $\tau=15$ ms, $\tau_v=600$ ms, $\sigma_R=5$ cm, $\eta_J=5\cdot10^{-4}$ Hz, $m=0.2$, $A_R=30 \,\, \text{Hz}$, $\alpha_J=1$, $v=26.8\,\,\text{cm/s}$, where syn denotes an arbitrary unit of synaptic weight. \textbf C, As \textbf B, but for learned weight amplitude, using the same parameters. Simulation results are summarized by the median and interquartile ranges across neurons, with the dashed line indicating the theoretical prediction (Eq.~\ref{eq:AJ-Cbeta}). \textbf D, Evolution of the feedforward connectivity matrix over time. Random initial weights gradually converge to a smooth Gaussian profile, enabling the competitive layer to develop a continuous spatial code. }
    \label{fig:SelfOrganizedWeights}
\end{figure*}

Simulations of a network without recurrent connectivity show that a translationally invariant input, combined with Hebbian learning, naturally produces a Gaussian profile in the feedforward weights. Fig.~\ref{fig:SelfOrganizedWeights}A illustrates the resulting system configuration. Global inhibition induces competition among neurons, Hebbian plasticity strengthens connections of consistently active neurons, while synaptic depression and adaptation prevent runaway activity. Without adaptation, a single neuron could dominate and learn all stimuli; with adaptation, suppression ensures that activity is distributed across the population. 

To study the steady states analytically (see Appendix \ref{app:MathematicalBasis} for derivations), we assume Gaussian-shaped equilibrium profiles for the activity variables (see also \cite{mi2014spike}):

\begin{align}
    U^{eq}(x,z) &= A_u\,\mathcal N(x; \,\, z,\sigma_u),\label{eq:anzatz-U} \\ 
    V^{eq}(x,z) &= A_v\,\mathcal N(x; \,\, z-d,\sigma_u),\label{eq:anzatz-V} \\
    r^{eq}(x,z) &= A_r\,\mathcal N\!\left(x; \,\, z,\frac{\sigma_u}{\sqrt2}\right),\label{eq:anzatz-r}
\end{align}

\noindent where $z=vt$ is the position of the moving stimulus. Here, $A_u$, $A_v$, and $A_r$ are the amplitudes of the respective variables, $\sigma_u$ is the bump width, and $d$ represents the lag of the adaptation variable $V(x,t)$ behind the moving bump, reflecting the delayed effect of adaptation on network activity. All amplitudes and $d$ depend on system parameters (details in Appendix \ref{app:MathematicalBasis}).

Since learning is much slower than bump dynamics, Hebbian updates can be averaged over a full stimulus cycle, which is equivalent to averaging Eq.~\ref{eq:hebbian-learning} over all possible positions of the bump. By setting the averaged time derivative to zero, the equilibrium condition yields

\begin{equation}
    J^{eq}(x,x^{\prime}) = 
    \left(
    \frac{\langle r^{eq}(x^{\prime},z)R(x,z)\rangle_z}{\alpha_J \langle r^{eq}(x^{\prime},z)\rangle_z}
    \right)^{1/\beta},
    \label{eq:equilibrium-cond-J}
\end{equation}

\noindent where $\langle \cdot \rangle_z$ denotes the average over all possible $z$ values.

Because the equilibrium activity profiles $r(x^{\prime},z)$ and $R(x,z)$ are Gaussian functions, Eq.~\ref{eq:equilibrium-cond-J} admits a Gaussian solution. Evaluating the averages results in

\begin{equation}
        J^{eq}(x,x^{\prime}) = A_J\,\mathcal N(x;\,\, x^{\prime},\sigma_J),\label{eq:anzatz-J}
\end{equation}

\noindent where

\begin{equation}
    A_J = \left(\frac{A_R}{\alpha_J C_\beta}\right)^{1/\beta}, \qquad 
        C_\beta = \sqrt{\frac{(2\pi\sigma_J^2)^{1-\beta}}{\beta}}.\label{eq:AJ-Cbeta}
\end{equation}

For self-consistency, all Gaussian terms in Eqs.~\ref{eq:U-dynamics},~\ref{eq:V-dynamics}~and~\ref{eq:equilibrium-cond-J} must share the same width. This requirement leads to the relations
\begin{equation}
    \sigma_J = \sqrt{\frac{3\beta}{2-\beta}}\,\sigma_R,
    \qquad
    \sigma_u = \sqrt{\frac{2\beta+2}{2-\beta}}\,\sigma_R.
    \label{eq:consistent-sigmas}
\end{equation}

These expressions establish a monotonic dependence of the learned widths on the parameter $\beta$, while also constraining its admissible range. Requiring real-valued solutions implies $0<\beta<2$, and the limit $\beta\to 2$ leads to a divergence of both widths. In particular, $\beta=0.5$ corresponds to the critical value for which the learned neural representation, in terms of the firing rate $r(x,z)$, has the same width as the input, i.e., $\sigma_u/\sqrt{2}=\sigma_R$. For $\beta<0.5$, the neural code is narrower than the input, yielding a more spatially precise representation. For $\beta>0.5$, the neural code becomes broader and therefore less spatially precise.

Starting from random initial weights, we simulated the system across a range of $\beta$ values. As shown in Figs.~\ref{fig:SelfOrganizedWeights}B-C, the learned Gaussian connectivity profiles recover an amplitude and standard deviation that closely align with the theoretical predictions of Eqs.~\ref{eq:AJ-Cbeta}~and~\ref{eq:consistent-sigmas}. Fig.~\ref{fig:SelfOrganizedWeights}D presents snapshots of an example connectivity matrix at various stages of the simulation, illustrating its temporal evolution towards the predicted Gaussian profile. Together, the alignment between these analytical derivations and computational simulations shows that the network robustly self-organizes from random initial conditions into a stable, predictable spatial representation.

\subsubsection{Weight stability}

We next study the stability of the connectivity profiles by means of a perturbative approach. In the regime of extremely weak inhibition ($k\ll 1$), small perturbations of the synaptic weights, $\delta J(x,x^{\prime})$, obey the linearized dynamics

\begin{equation}
    \begin{aligned}
    \tau_J(x,x^{\prime})&\frac{\partial}{\partial t}\delta J(x,x^{\prime}) \\
    =&\int dx^{\prime\prime}  K(x,x^{\prime},x^{\prime\prime})\delta J(x^{\prime\prime},x^{\prime}) - \delta J(x,x^{\prime}),
    \end{aligned}\label{time dep diff eq}
\end{equation}

\noindent where $K(x,x^{\prime},x^{\prime\prime})$ denotes the effective interaction kernel and $\tau_J(x,x^{\prime})$ is an effective, spatially dependent time constant (see Appendix \ref{app:FFWeightStability} for explicit expressions and derivation). We note that both $J^{eq}(x,x^{\prime})$ and $K(x,x^{\prime},x^{\prime\prime})$ depend on the relative coordinates. This allows rewriting the dynamics in terms of the displacement variables $s=x-x^{\prime}$ and $s^{\prime\prime}=x^{\prime\prime}-x^{\prime}$ and a reparametrized perturbation $w_{x^{\prime}}(s,t)=\delta J(s+x^{\prime},x^{\prime},t)$ (see Appendix \ref{app:FFWeightStability}). Under this transformation, the dynamics reduce to an operator equation of the form

\begin{equation}
    \tau_J(s)\frac{\partial}{\partial t}w_{x^{\prime}}(s,t) = \hat L\left[w_{x^{\prime}}(s,t)\right],
\end{equation}

\noindent where $\hat L$ is a linear integral operator defined by

\begin{equation}
    \hat L\left[w_{x^{\prime}}(s,t)\right] = \int ds^{\prime\prime} K(s,s^{\prime\prime})w_{x^{\prime}}(s^{\prime\prime},t) - w_{x^{\prime}}(s,t).
\end{equation}

We then seek the modal solutions $w_{x^{\prime}}(s,t)=\sum_ic_i(x^{\prime})f_i(s)e^{\lambda_i t}$, which lead to the generalized eigenvalue-eigenfunction problem

\begin{equation}
    \hat L\left[f_i(s)\right] = \lambda_i\tau_J(s)f_i(s).
\end{equation}

\begin{figure}[h!]
    \centering
    \includegraphics[width=\linewidth]{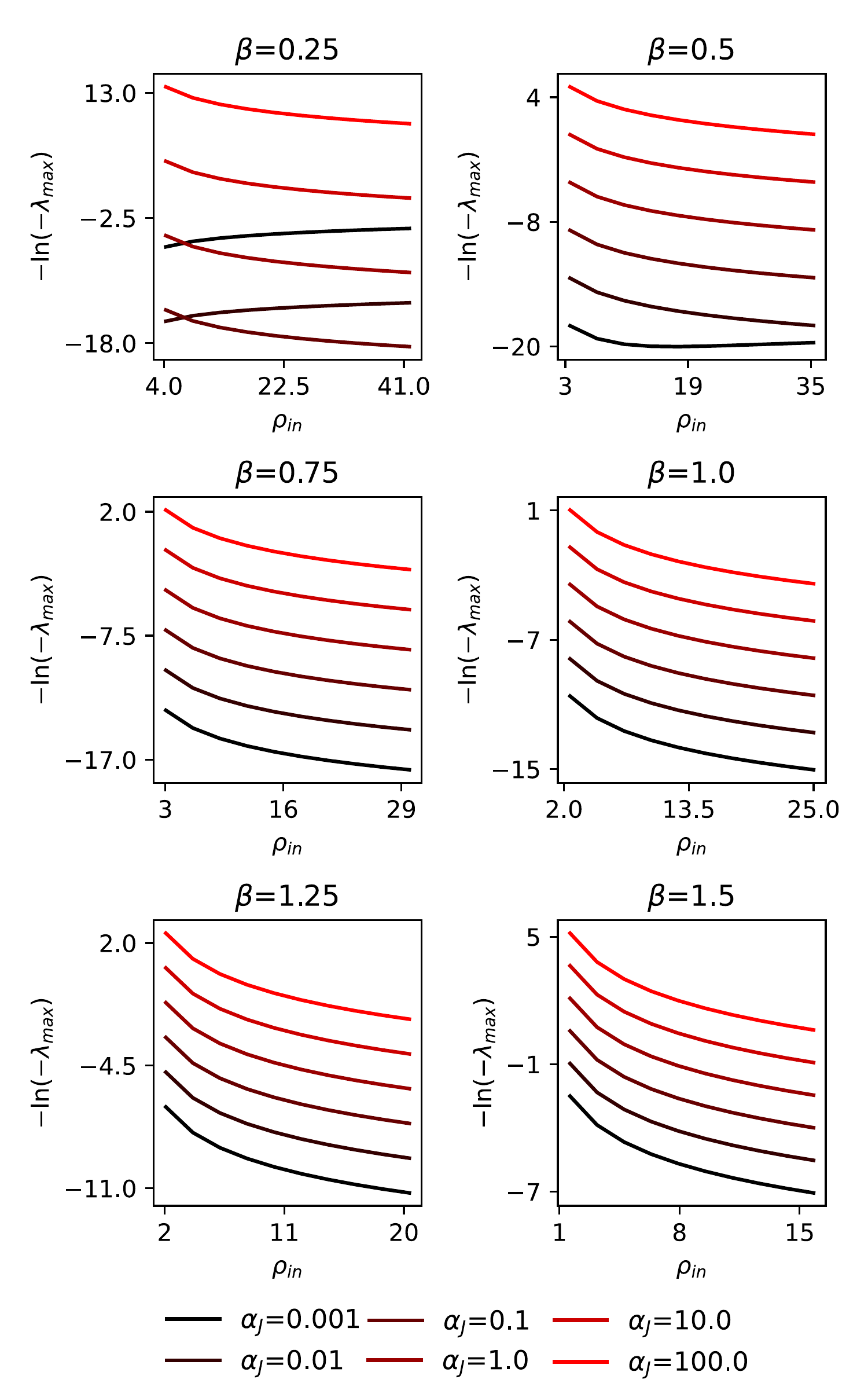}
    \caption{\textbf{Global stability analysis across inhibition factor and coupling strength}. Numerical spectral analysis of the generalized eigenvalue problem shows the dependence of the real part of the leading eigenvalue, $\lambda_\text{max}$, on the input density $\rho_\text{in}$ and synaptic damping strength $\alpha_J$, for different values of $\beta$. The system is stable whenever $\Re(\lambda_i)<0$ for all modes; accordingly, the curves report $-\ln(-\lambda_\text{max}$). Across the explored parameter range, $\lambda_\text{max}$ remains negative, indicating robust stability. Panels correspond to the specified $\beta$ values. Colors indicate $\alpha_J$ from 0.001 to 100. Spectral analysis was performed with a spatial resolution of $\Delta s=0.1\,\,\text{cm}$ and system parameters $m=0.15$, $\tau=15\,\, \text{ms}$, $\tau_v=600 \,\,\text{ms}$, $k=0.01$ $\text{cm}^4\text{s}^2/\text{syn}$, $\eta_J= 5 \cdot 10^{-4}\,\,\text{Hz}$, $A_R=30\,\,\text{Hz}$, $\sigma_R =5\,\,\text{cm}$, $L=20\sqrt{\frac{2\beta+2}{2-\beta}}\sigma_R$, and $\rho_\text{in} =N_\text{in}/L$, with $N_\text{in}$ ranging from 512 to 5000 neurons (across the tested parameter range, no eigenvalues with positive real parts were observed).}
    \label{fig:TheoreticalStabilityLambdas}
\end{figure}

The system can only be stable if all the eigenvalues satisfy $\Re(\lambda_i)<0$. Importantly, the spectrum only depends on the displacement variables (not $x^{\prime}$), so this condition defines a global stability criterion (see Appendix \ref{app:FFWeightStability} for details on the spectral decomposition). Numerical spectral analysis reveals that the eigenvalue with highest real part (from now on, referred as $\lambda_\text{max}$) remains negative across a broad parameter range, including biologically plausible parameters. Specifically, we show in Fig.~\ref{fig:TheoreticalStabilityLambdas} the change in $\lambda_\text{max}$ across a wide range of input densities ($\rho_{in}$) and damping synaptic strengths ($\alpha_J$), focusing on these because other biologically relevant parameters did not yield such prominent variations.

Our theoretical analysis complements the simulation results. In the biologically plausible parameter regime (and beyond), the system remains theoretically stable as long as the inhibition is sufficiently weak. However, it is of biological interest to determine what happens in a higher inhibition regime where analytical manipulation becomes challenging. Our simulations demonstrate that increasing the inhibition factor $k$ does not lead to instabilities. In fact, all simulation results presented in this work involving learned feedforward connectivity correspond to simulations with non-negligible values of $k$. Thus, both approaches converge on a single conclusion: the learned feedforward weights are fundamentally stable, ensuring a robust spatial representation across a wide spectrum of network parameters.

\subsection{\label{subsec:PredictiveCoding}Bump Asymmetry and Anticipative Coding}

In the preceding sections we have shown that Gaussian connectivity profiles and their resulting population activity profiles constitute stable solutions when driven by translationally invariant moving inputs. This self-organizing property facilitates the construction of hierarchical feedforward architectures, wherein each successive layer learns to represent the manifold provided by its predecessor. We now investigate the structural distortions of the population bump profile as a function of the network’s intrinsic parameters.

From a functional perspective, these distortions are naturally characterized by an expansion in Hermite polynomials (see Appendix \ref{app:MathematicalBasis} for a derivation, or \cite{fung2008dynamics} for a comprehensive treatment). In this framework, the Gaussian profile corresponds to the zeroth-order Hermite mode, while deviations from symmetry are captured by higher-order terms. The odd-order modes are of particular interest as they govern the displacement of the bump’s center of mass. Specifically, our analysis reveals that increasing the adaptation strength systematically enhances the contribution of the first-order mode, inducing a pronounced asymmetry in the activity profile.

To quantify this effect, we define an ansatz for the steady-state moving bump solution:

\begin{equation}
    U^{eq}_{\gamma}(x,z) = A_{u}\, \mathcal{N}(x;\,\, z, \sigma_u) \left(1 + \gamma \frac{x-z}{\sigma_u}\right),
    \label{eq:U-ansatz-1}
\end{equation}
\noindent where $\gamma$ scales the contribution of the first-order Hermite polynomial and serves as the primary metric for profile asymmetry. For small perturbations ($\gamma\ll 1$), this expansion is equivalent to a first-order Taylor approximation of a shifted Gaussian:

\begin{equation}
    \mathcal N (x;\,\, z+\sigma_u \gamma, \sigma_u) \approx \mathcal N (x;\,\, z, \sigma_u)\left(1+\gamma\frac{x-z}{\sigma_u}\right) .
    \label{eq:U-approx-shift}
\end{equation}

This approximation illustrates that a positive $\gamma$ corresponds to a forward displacement of the activity peak, effectively encoding a predictive representation of the trajectory. Conversely, a negative $\gamma$ implies retrospective coding. Thus, the emergence of a non-zero first-order Hermite coefficient provides a formal mathematical signature of either prospective or retrospective neural dynamics.

By substituting the ansatz from Eq.~\ref{eq:U-ansatz-1} into the governing dynamical equations (Eqs.~\ref{eq:U-dynamics}~and~\ref{eq:V-dynamics}) and projecting the system onto the first two Hermite modes, we derive an explicit relationship between the asymmetry parameter $\gamma$ and the spatial lag $d$ between the excitatory variable $U(x,t)$ and the adaptive variable $V(x,t)$ (detailed in Appendix \ref{app:EmergentAsym}). To facilitate the analysis, we introduce the following dimensionless variables:

\begin{equation}
    u = \frac{\tau v}{\sqrt{2}\sigma_u}, \quad y = \frac{d}{\sqrt{2}\sigma_u}, \quad \Gamma = \frac{\tau_v}{\tau}.
\end{equation}

The steady-state solutions for the asymmetry $\gamma$ and the spatial displacement $y$ are given by: 

\begin{equation}
    \gamma = \frac{my}{\Gamma y u + 1} - u,
\end{equation}
and
\begin{equation}
    y = \frac{(m+1) - \Gamma u^2}{2\Gamma u} \left( \sqrt{1 + \frac{4u^2 \Gamma (\Gamma + 1)}{\left[ \Gamma u^2 - (m+1) \right]^2}} - 1 \right).
    \label{eq:gamma-y-expressions}
\end{equation}

\begin{figure}[h!]
    \centering
    \includegraphics[width=1\linewidth]{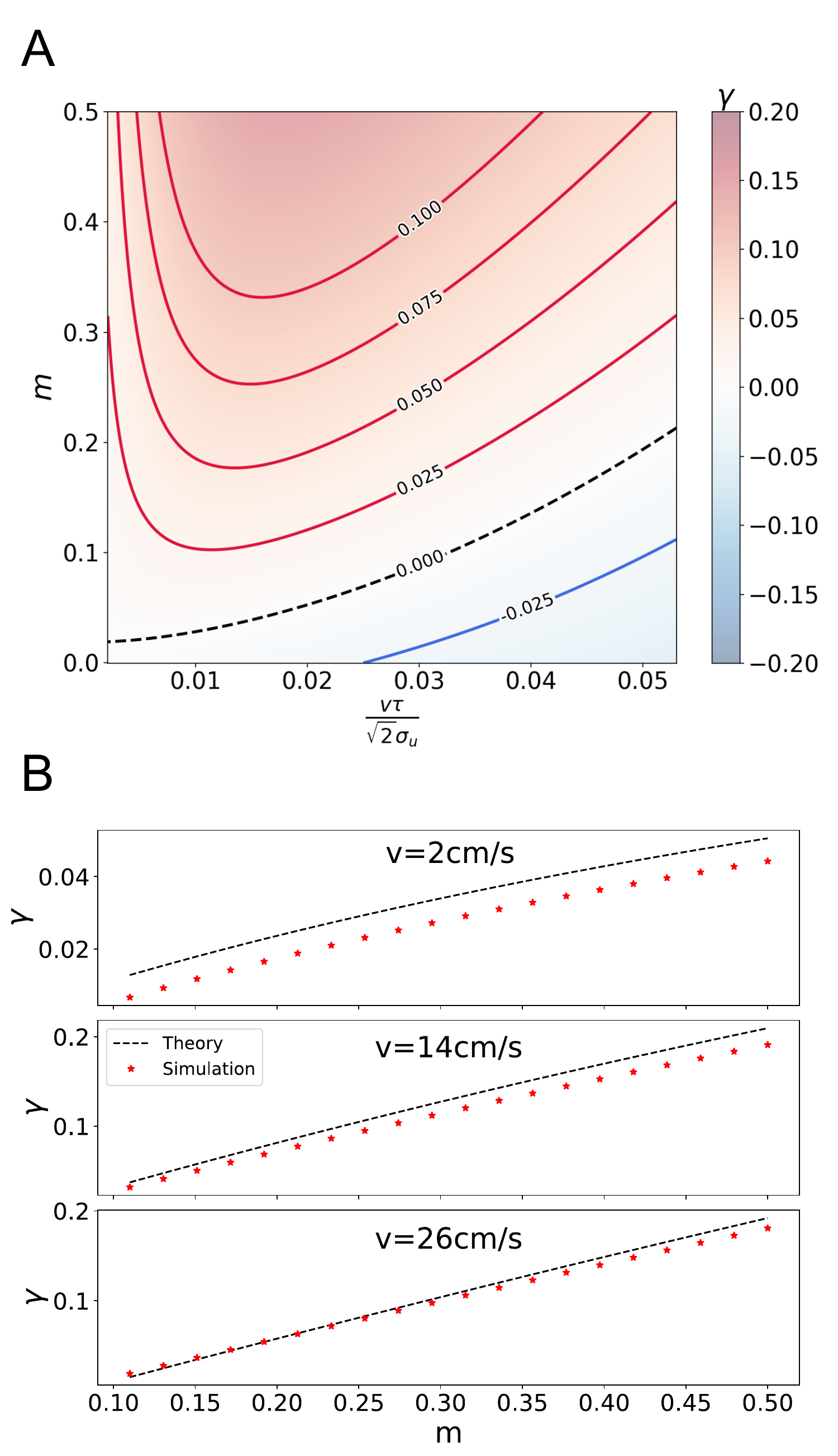}
    \caption{\textbf{Dependence of bump asymmetry \(\gamma\) on adaptation strength and input speed}.
        \textbf A, Level curves of constant \(\gamma\) in the $(m, \tfrac{\tau v}{\sqrt2\sigma_u})$ parameter space, showing how combinations of adaptation strength $m$ and input velocity $v$ give rise to different bump asymmetries.
        \textbf B, Theoretical predictions (dashed black lines) and simulation results (red stars) for $\gamma$ as a function of $m$, for three fixed values of input velocity $v$. Each curve corresponds to a different $v$, illustrating the modulation of bump asymmetry by the speed of external input as well as its monotonic dependence on $m$. Simulations were performed using the Euler method with a time step of $\Delta t=5\cdot 10^{-3}$ s. Parameters used: $N_\text{in}=N_c=1000$, $k = 0.1$ $\text{cm}^4\text{s}^2/\text{syn}$, $\tau = 10$ ms, $\tau_v = 800$ ms, $\sigma_R = 5$ cm, $A_R = 30\,\,\text{Hz}$.
        }
    \label{gamma_theoretical_and_sim}
\end{figure}

Fig.~\ref{gamma_theoretical_and_sim} shows that the equilibrium value of $\gamma$ increases with adaptation strength $m$, with the exact relationship modulated by the input speed $v$. This demonstrates that adaptation introduces a systematic asymmetry into the bump profile. Importantly, the sign and magnitude of the asymmetry are not dictated by adaptation alone, but emerge from the interaction between adaptation and input velocity. Larger $m$ accelerates the decay of neural activity, and through global inhibition this decay allows previously silent neurons to become active, shifting the bump in the direction of motion. If activity decays quickly enough, the network can anticipate the stimulus (positive asymmetry). Conversely, weak adaptation or very high input velocity prevents sufficient recruitment of new neurons, resulting in delayed responses and retrospective coding (negative asymmetry).

\begin{figure*}[t]
    \centering
    \includegraphics[width=1\linewidth]{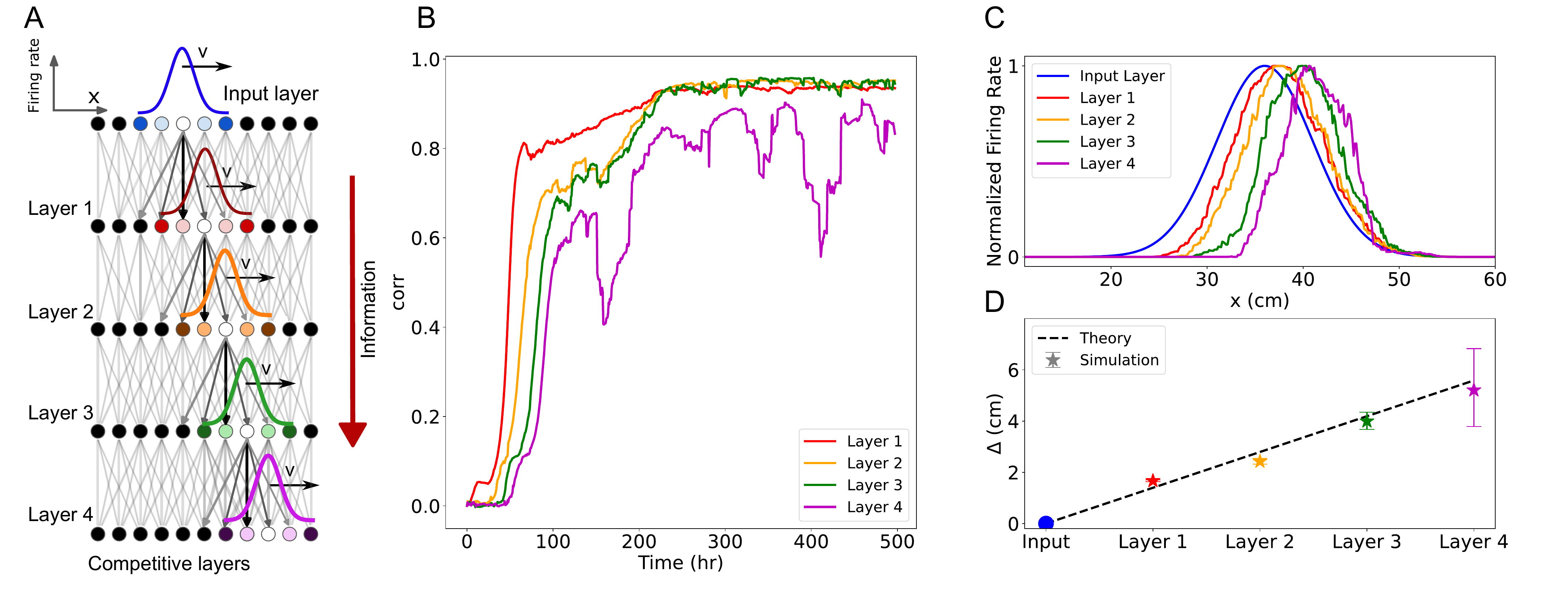}
    \caption{\textbf{Self-organized multilayer architecture potentiates predictive coding}.
        \textbf{A}, Schematic of the multilayer model after learning. Each layer successfully learns to encode space. Additionally, each layer encodes a position that is shifted into the future relative to its preceding input layer. \textbf{B}, The learning process for each of the four layers. The plot shows the correlation between each layer’s feedforward connectivity matrix and the theoretically expected Gaussian profile. Despite being initialized with random weights, all layers converge to a high correlation, indicating that they all learn to encode space. \textbf{C}, An example of the activity bump in each layer compared to the original input signal. The spatial shift, or prediction into the future, increases progressively with the layer number. \textbf{D}, The predictive shift as a function of the layer number. The model’s theoretical prediction $\Delta_i = L_i \gamma\sigma_u$, with $L_i$ the number of layer (dashed black line). The shift is computed as the difference between the center of mass of the layer's bump and that of the input signal. Stars denote the median shift computed over 100 positions uniformly sampled in $[0,L)$, with error bars indicating the corresponding interquartile range.  All shown results correspond to simulations performed using the Euler method with a time step of $\Delta t=5\cdot 10^{-3}$ s and done for the set of parameters $\tau=15$ ms, $\tau_v=600$ ms, $m=0.25$, $\sigma_R=5$ cm, $A_R=30\,\,\text{Hz}$, $\eta_J=5\cdot10^{-4}$ Hz, $\alpha_J=0.5$, $k=0.01$ $\text{cm}^4\text{s}^2/\text{syn}$, $L=70$ cm, and $v=20$ cm/s, $N_\text{in}=N_c=512$ for all layers (including input). 
}
    \label{fig:multilayer}
\end{figure*}

In a feedforward architecture composed of multiple stacked layers, each successive layer not only reconstructs the underlying one-dimensional manifold but also displays an activity bump shifted forward with respect to its predecessor (see schematic in Fig.~\ref{fig:multilayer}A). Importantly, these predictive shifts do not strictly rely on a deep architecture; similar anticipatory behavior can be achieved within a single layer by tuning the adaptation strength. Instead, the primary role of a multilayer hierarchy is to provide a robust framework through which these shifts can naturally emerge and be progressively potentiated through learning. Fig.~\ref{fig:multilayer}B illustrates this evolutionary process, tracking how every layer within the network progressively shifts from completely random initializations toward structured profiles by measuring the correlation between each connectivity matrix and its theoretically expected Gaussian shape.

Analytically, this progressive potentiation means that for a system of $M$ layers, the final layer is expected to display a cumulative forward shift given by $\Delta = M\sigma_u\gamma$ (see Figs.~\ref{fig:multilayer}C~and~D). However, this linear amplification comes at the cost of error accumulation across successive processing stages. This compounding error imposes a natural upper bound on the viable depth of the feedforward network, as sequential information degradation eventually compromises the accuracy and stability of the spatial prediction.

This provides a mechanistic interpretation of how anticipatory codes may arise naturally through learning in dynamic environments. Such predictive shifts are consistent with experimental observations in the medial entorhinal cortex, where neurons in superficial layers exhibit activity that anticipates the animal’s future trajectory \citep{kropff2015speed,ouchi2024predictive}. Our results suggest that hierarchical feedforward structures can be learned through self-organizing rules, enabling anticipatory coding without requiring strong recurrent interactions. This is consistent with the observation that a relatively low density of recurrent connections is found in superficial layers of the entorhinal cortex \citep{dhillon2000laminar,couey2013recurrent,pastoll2013feedback}, precisely where anticipatory activity is most prominent.

\subsection{\label{subsec:RecurrentWeights}Learning Recurrent Weights}

\subsubsection{Stability Analysis}

\begin{figure*}[t]
    \centering
    \includegraphics[width=\textwidth]{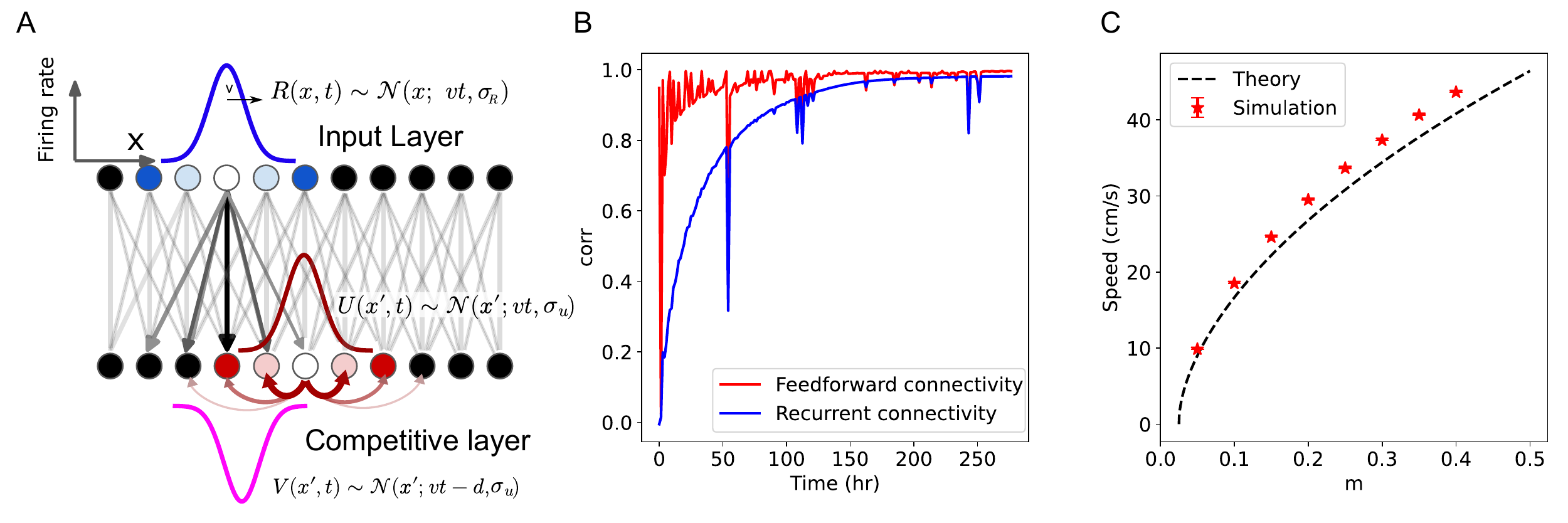}
    \caption{\textbf{Recurrent attractor connectivity can be learned from an input tutor}. \textbf A, Schematic of the two layer network after learning both feedforward and recurrent connectivity. \textbf B, The learning process for both recurrent and feedforward connectivities. The plot shows the correlation between each connectivity matrix and the theoretically expected Gaussian profile. \textbf C, The learned recurrent connectivities allow the network to self-sustain a moving Gaussian bump in the absence of input tutor. The speed of the Gaussian bump satisfies the theoretical prediction Eq.~\ref{eq:theovel}. Stars denote the median speed computed over $10^4$ simulation steps (Euler integration, $\Delta t=5\cdot 10^{-3}$ s), with error bars indicating the corresponding interquartile range. Parameters used: $N_\text{in} = N_c=512$, $L=100$ cm, $k=5.12\cdot 10^{-1}$ $\text{cm}^4\text{s}^2/\text{syn}$, $\tau=15$ ms, $\tau_v=600$ ms, $\sigma_R=5$ cm, $\eta_J=\eta_W=5\cdot10^{-4}$ Hz, $A_R=30$ Hz, $\alpha_J=1$, $\alpha_W=8$, $\beta=0.5$ and $v$ computed from Eq.\ref{eq:theovel}. }
    \label{fig:RecurrentWeights}
\end{figure*}

In this section, we incorporate recurrent connectivity into the proposed framework (see schematic in Fig.~\ref{fig:RecurrentWeights}A). While the feedforward architecture described above accounts for anticipatory shifts, it does not provide a mechanism for path integration. Estimating position from self-motion requires the continuous integration of proprioceptive signals, such as velocity and heading direction, which in neural systems is typically implemented through recurrent dynamics. In particular, CANN models rely on structured recurrent connectivity to translate velocity inputs into coherent shifts of the activity bump along the manifold.

Accordingly, we extend the model by introducing a recurrent contribution to the synaptic current. The total synaptic current $I(x,t)$ is reformulated as the sum of feedforward and recurrent contributions:

\begin{equation}
    \begin{aligned}
    I(x,t) =& \underbrace{\rho_{\text{in}}\int_{x^{\prime}}dx^{\prime} J(x,x^{\prime})R(x^{\prime},t)}_{I_\text{ff}(x,t)} \\ &+ \underbrace{\rho_c\int_{x^{\prime}}dx^{\prime}W(x,x^{\prime})r(x^{\prime},t)}_{I_\text{rec}(x,t)},
    \end{aligned}
\end{equation}

\noindent where $W(x,x^{\prime})$ represents the recurrent connectivity matrix. The temporal evolution of $W(x,x^{\prime})$ is governed by a Hebbian plasticity rule analogous to the one defining the feedforward weights $J(x,x^{\prime})$ (Eq.~\ref{eq:hebbian-learning}):

\begin{equation}
\frac{\partial }{\partial t} W(x,x^{\prime}) = \eta_W r(x,t)\left[r(x^{\prime},t) - \alpha_W W(x,x^{\prime})^\beta \right].\label{eq:learning-rule-recurrent}
\end{equation}

Under the conditions specified in this work (where spatial inputs come from Gaussian tuning curves in the tutor layer), this learning rule gives rise to a Gaussian connectivity profile that we will express as $W^{eq}(x,x')=A_W\mathcal{N}(x;\,\, x^{\prime}, \sigma_W)$. To ensure structural consistency across the network, we focus on the case $\beta=0.5$. This specific parameterization allows the system to converge toward feedforward and recurrent connectivity matrices with matched standard deviations, thereby harmonizing the spatial scales of both input sources (see Eq.~\ref{eq:consistent-sigmas} and Fig.~\ref{fig:SelfOrganizedWeights}B).

Numerical simulations show that the system can learn structured feedforward and recurrent connectivity. Fig.~\ref{fig:RecurrentWeights}B  shows the evolution of the correlation between each learned connectivity matrix and the theoretical prediction. By arguments analogous to the feedforward case, the expected recurrent connectivity is approximately Gaussian with standard deviation equal to $\sigma_R$ (for $\beta=0.5$). 

After learning the attractor recurrent connectivity, the system is capable of self-sustaining a moving bump of activity with intrinsic speed $v_\text{int}(m)$ monotonically dependent on the adaptation strength \cite{mi2014spike}: 

\begin{equation}
    v_\text{int}(m) = \frac{\sqrt2\sigma_u}{\tau_v}\sqrt{\frac{m\tau_v}{\tau} - \sqrt{\frac{m\tau_v}{\tau}}}
    \label{eq:theovel}
\end{equation}

In this and subsequent sections, we present theoretical results corresponding to a corrected version of Eq.~\ref{eq:theovel}. We rescale the theoretical formula by an empirical factor of 0.72 to account for inaccuracies introduced by the projection method's Gaussian ansatz. This correction factor was derived from simulations across various combinations of $\tau$ and $\tau_v$ (see Appendix \ref{subsubapp:ProjMethUPI}). Fig.~\ref{fig:RecurrentWeights}C shows that the resulting bump velocities are in good agreement with the corrected theoretical prediction.

\begin{figure*}[t]
    \centering
    \includegraphics[width=\textwidth]{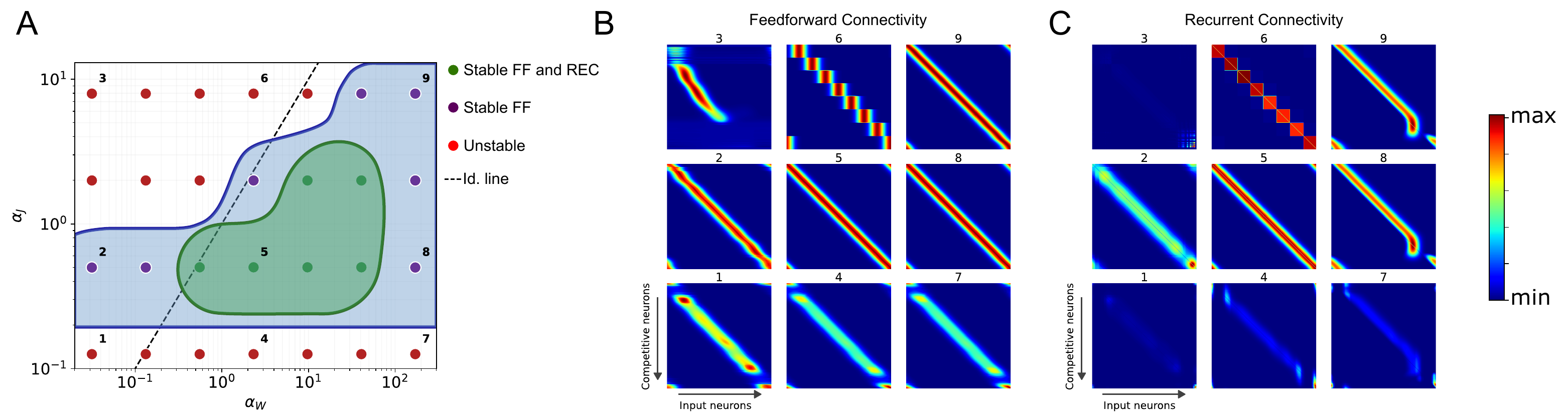}
    \caption{\textbf{Stability landscape of recurrent connectivity in the $(\alpha_J, \alpha_W)$ parameter space}. This figure summarizes how hyperparameter tuning shapes the emergence of recurrent attractor structure in the network. \textbf A, Regions of the parameter space in which the learned feedforward (FF) and recurrent (REC) connectivities achieve high correlation with their target Gaussian profiles. Red dots indicate parameter combinations for which neither feedforward nor recurrent connectivity attains a high final correlation. Blue dots indicate cases in which the feedforward connectivity converges to a high-correlation solution, whereas the recurrent connectivity does not. Green dots indicate parameter combinations in which both feedforward and recurrent connectivities converge to high-correlation Gaussian profiles, enabling the system to self-sustain a moving bump of activity. The blue shaded region schematizes the parameter regime in which learned feedforward connectivity is expected, whereas the green shaded region schematizes the regime in which fully stable attractor dynamics are expected. \textbf B, Representative examples of the final feedforward connectivity matrices for the parameter combinations indicated in Panel \textbf A. \textbf C, Representative examples of the final recurrent connectivity matrices for the parameter combinations indicated in Panel \textbf A. In both panels \textbf B and \textbf C, numbers above each matrix correspond to the associated point in the phase diagram. Simulations were performed using the Euler method with a time step of $\Delta t=5\cdot 10^{-3}$ s. Simulation parameters: $N_\text{in}=N_c=512$, $\tau=15$ ms, $\tau_v=600$ ms, $m=0.2$, $\sigma_R=5$ cm, $A_R=30$ Hz, $\eta_J=\eta_W=10^{-3}$ Hz, $k=5.12\cdot10^{-1}$ $\text{cm}^4\text{s}^2/\text{syn}$, $L=100$ cm, $\beta=0.5$ and $v=26.8$ cm/s.}
    \label{fig:RecurrentStability}
\end{figure*}

Regarding stability, the complexity of the coupled dynamics makes an analytical treatment challenging. We therefore resort to extensive numerical simulations to identify the parameter region in which excitatory recurrent connectivity can emerge and remain stable. By performing a grid search in the $(\alpha_J,\alpha_W)$ plane, we find that $\alpha_W \gtrsim \alpha_J$ is required for the system to effectively learn the feedforward connectivity. Since $\alpha_J$ is inversely related to the feedforward weight amplitude $A_J$, and $\alpha_W$ to $A_W$ (see Eq.~\ref{eq:AJ-Cbeta}), this result indicates that the regime in which the system learns to encode space requires the tutor-driven input to dominate over the recurrent signal (see Fig.~\ref{fig:RecurrentStability}B). Interestingly, recurrent weights can be learned provided that the recurrent signal is not too weak (see Fig.~\ref{fig:RecurrentStability}C). Moreover, the system fails to maintain stable feedforward connectivity for sufficiently small $\alpha_J$ (see Fig.~\ref{fig:RecurrentStability}B). A plausible explanation for this behavior follows from the approximation underlying the stability analysis: learning must remain slow. If the input current becomes too strong, the time derivative of the weights can increase to the point where the slow-learning-by-averaging condition is violated. Overall, while extreme input strengths compromise stability, the identified optimal regime clearly shows that a network can simultaneously self-organize its feedforward and recurrent architectures, successfully giving rise to self-sustained continuous attractor dynamics.

\subsection{\label{subsec:UPI}Unidirectional Path Integration}

Finally, we assess whether the competitive neural layer can function as a self-organized, unidirectional path integrator. In biological systems, velocity is time-varying; in contrast, the emergent attractor in our model propagates at a speed determined by intrinsic parameters—adaptation strength ($m$), bump width ($\sigma_u$), and the characteristic time constants ($\tau$, $\tau_v$). Consequently, the network cannot support accurate path integration under variable velocity conditions.

To overcome this limitation, we introduce an external control signal in the form of a spatially uniform background current that encodes the instantaneous speed—an approach similar in spirit to the velocity inputs utilized in established path integration models (see \citep{burak2009accurate} and \citep{widloski2014model}):

\begin{equation}
    I(x,t) = I_\text{ff}(x,t) + I_\text{rec}(x,t) + \underbrace{g [v(t) - v_\text{int}(m)]}_{I_\text{speed}},
\end{equation}

\noindent where $g$ is a positive gain factor and $v(t)$ is the external speed (which can vary dynamically in time).

In what follows, we exploit the analytical tractability of the model to provide a mechanistic account of how this input rescales the effective self-sustained speed of the bump of the attractor, thereby enabling controlled modulation of its propagation speed.

We analyze the system’s intrinsic dynamics in the absence of spatial feedforward input ($I_\text{ff}=0$). To derive a theoretical understanding of the motion, we employ the projection method. Unless otherwise noted, for sake of simplicity we assume a constant external speed $v$. We later show through numerical simulations that our results generalize for varying speeds. We introduce a modified ansatz that incorporates the effect of the global input current by allowing for a uniform baseline shift. Concretely, we propose that the solutions retain the shape of the original equilibrium profiles, but are offset by a constant term that depends on the input speed:

\begin{equation}
    U^{eq}_\epsilon(x,z) = U^{eq}(x,z) + \epsilon_u I_\text{speed},\label{eq:anzatz-U-beta}
\end{equation}

\begin{equation}
    V^{eq}_\epsilon(x,z) =  V^{eq}(x,z) + \epsilon_v I_\text{speed},\label{eq:anzatz-V-beta}
\end{equation}

\noindent where $U^{eq}(x,z)$ and $V^{eq}(x,z)$ denote equilibrium solutions of the form defined in Eqs.~\ref{eq:anzatz-U}~and~\ref{eq:anzatz-V}. The scaling factors $\epsilon_u$ and $\epsilon_v$ are constrained by the steady-state conditions of the network dynamics:

\begin{equation}
    \epsilon_u + \epsilon_v = 1, \quad \epsilon_v = m\epsilon_u.
\end{equation}

It is important to note that substituting this ansatz into the dynamical equations yields a system effectively identical to the standard recurrent network, as the constant terms vanish. The equations of motion for the fluctuating equilibrium components become:

\begin{equation}
    \begin{aligned}
        \tau\frac{\partial}{\partial t}U_\epsilon^{eq}(x,z(t)) =& -U^{eq}(x,z(t)) - V^{eq}(x,z(t)) \\ 
        &+ \rho_{c}\int_{x^{\prime}}  dx^{\prime} \, W^{eq}(x,x^{\prime})r_\epsilon^{eq}(x^{\prime},z(t)), 
    \end{aligned}
\end{equation}
\begin{equation}
    \begin{aligned}
        \tau_v\frac{\partial}{\partial t}V_\epsilon^{eq}(x,z(t)) &= - V^{eq}(x,z(t)) + mU^{eq}(x,z(t)).
    \end{aligned}
\end{equation}

However, a crucial difference arises from the nonlinearity of the firing-rate function. The global current alters the effective activation width of the network. Assuming the baseline shift is small relative to the bump amplitude ($\epsilon_uI_\text{speed}\ll\frac{A_u}{\sqrt{2\pi}\sigma_u}$), the firing rate can be approximated as:
\begin{equation}
    r_\epsilon^{eq}(x,z) \approx \frac{U^{eq}(x,z)^2}{B} + 2\epsilon_u I_\text{speed}\frac{U^{eq}(x,z)}{B}.\label{eq:linearized_req}
\end{equation}

This expansion introduces an additional term in the dynamical equation for $U(x,z)$. Because this new contribution has a mismatched standard deviation, it drives the system toward a different equilibrium, specifically one with a larger standard deviation. A useful, albeit approximate, way to analyze the effect of the speed input current on the standard deviation is to assume a stationary Gaussian bump. Although the actual solution corresponds to a moving bump, the stationary approximation is expected to capture the essential dependence of the standard deviation on the input current. Under this assumption, the steady-state bump profile can be expressed as 

\begin{equation}
    \begin{aligned}
    U^{eq}(x,z)\approx& \rho_c\int dx^{\prime} W^{eq}(x,x^{\prime})r_\epsilon^{eq}(x^{\prime},z) \\
    \approx& \frac{\rho_cA_u^2A_W}{B2\sqrt\pi\sigma_u}\mathcal N(x;\,z,\sigma_u) \\ 
    &+\frac{2\epsilon_uI_\text{speed}}{B}\rho_cA_uA_W\mathcal N\left(x;\,z,\sqrt\frac{3}{2}\sigma_u\right)
    \end{aligned}
\end{equation}

Therefore, we can roughly estimate the standard deviation of the bump by computing the zeroth ($M_0$) and second ($M_2$) moments of $U^{eq}(x,z)$

\begin{equation}
    \begin{aligned}
    M_0 &= \frac{\rho_cA_u^2A_W}{B2\sqrt\pi\sigma_u}\left(1+\frac{4\epsilon_uI_\text{speed}}{A_u}\sqrt\pi\sigma_u\right)\\
    M_2 &= \frac{\rho_cA_u^2A_W}{B2\sqrt\pi\sigma_u}\sigma_u^2\left(1+\frac{6\epsilon_uI_\text{speed}}{A_u}\sqrt\pi\sigma_u\right)
    \end{aligned}
\end{equation}

Assuming $\epsilon_uI_\text{speed}\ll\frac{A_u}{\sqrt{2\pi}\sigma_u}$, and neglecting variations in $\frac{A_u}{\sqrt{2\pi}\sigma_u}$, we may then expand the expression to first order in the small parameter $\epsilon_uI_\text{speed}\frac{\sqrt{2\pi}\sigma_u}{A_u}$:

\begin{equation}
    \text{std}\left[U^{eq}(x,z)\right]_x = \sqrt{M_2/M_0}\approx \sigma_u\left(1+\frac{\sqrt{\pi}\sigma_u}{A_u}\epsilon_uI_\text{speed}\right)\label{eq:change-std}
\end{equation}

Since the propagation speed of the bump depends on the spatial width of the activity profile, this distortion directly modulates the velocity. While the adaptation dynamics remain structurally unchanged, the modified standard deviation leads to a new intrinsic speed:

\begin{equation}
    \frac{dz}{dt} \approx v_\text{int}(m) \left(1+\frac{\sqrt{\pi}\sigma_u}{A_u}\epsilon_uI_\text{speed}\right),
    \label{eq:new-speed}
\end{equation}

\noindent where $v_\text{int}(m)$, given by Eq.~\ref{eq:theovel}, denotes the intrinsic speed of the bump when the network receives no external speed input current.

\begin{figure}[h]
    \centering
    \includegraphics[width=1.\linewidth]{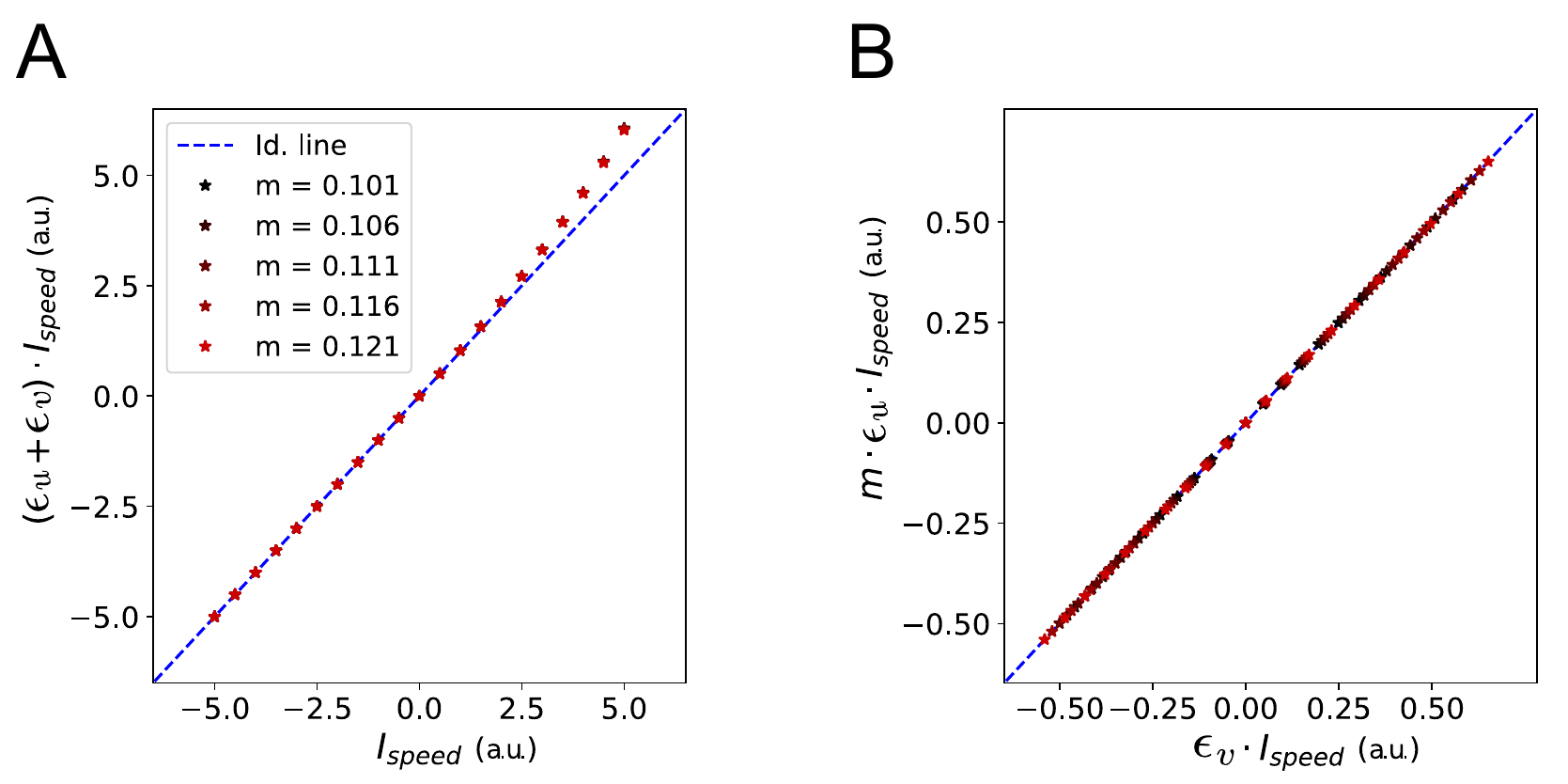}
    \caption{\textbf{Validation of the baseline shift properties}. \textbf A, Empirically measured baseline shifts in the $U(x,t)$ and $V(x,t)$ variables, demonstrating adherence to the conservation property. \textbf B, Measured baseline shifts showing consistency with the predicted proportionality property. A consistent color-coding scheme is maintained across both panels. Simulations were performed using the Euler method with a time step of $\Delta t=5\cdot 10^{-3}$ s. Parameters used for simulations: $N_c=1000$, $L=100$ cm, $A_W=1$ $\text{syn}$, $\tau=10$ ms, $\tau_v=100$ ms, $k=1.25\cdot 10^{-3}$ $\text{cm}^4\text{s}^2/\text{syn}$, $g=1$ $\text{cm}^{-3}$, $v_0=12.59$ cm/s.
}
    \label{fig:beta-constants}
\end{figure}

Numerical simulations validate this ansatz. As shown in Fig.~\ref{fig:beta-constants}, the system maintains stable, baseline-shifted Gaussian bumps, with the observed constants $\epsilon_u$ and $\epsilon_v$ satisfying the predicted constraints.

\begin{figure*}[t]
    \centering
    \includegraphics[width=\textwidth]{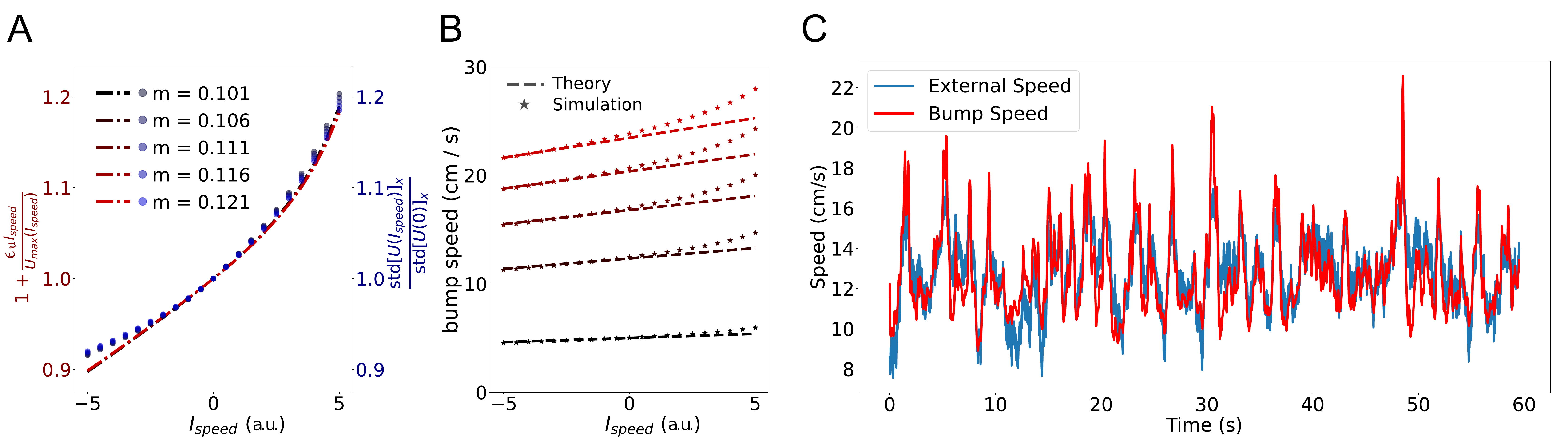}
    \caption{\textbf{Unidirectional path integration under constant and time-varying speed}. \textbf{A}, The right axis shows the measured standard deviation of the activity profile, $\mathrm{std}[U(I_\text{speed})]_x$, as a function of $I_\text{speed}$, normalized by its value at $I_\text{speed} = 0$. The left axis shows the measured baseline shift, $\epsilon_u I_\text{speed}$, normalized by the maximum value of the $U(x,t)$ bump ($U_{\max}$) for each different $I_\text{speed}$, theoretically equivalent to $\frac{A_u}{\sqrt{2\pi}\sigma_u}$. Both quantities exhibit close agreement with the dependence on $I_\text{speed}$, consistent with Eq.~\ref{eq:change-std}. The dashed lines are shown in a red scale and the dots in a blue scale to indicate different $m$ values. \textbf{B}, Measured bump velocities across varying adaptation levels compared against our theoretical prediction (Eq.~\ref{eq:new-speed}) as a function of the input current $I_\text{speed}$. The red color scale corresponds to the same set of $m$ values shown in panel \textbf A. \textbf{C}, Path integration performance under time-varying speed. The input current modulates the bump velocity in real time for $m=0.106$. The corresponding simulation parameters in all panels are the same as those in Fig.~\ref{fig:beta-constants}.}
    \label{fig:UPI-results}
\end{figure*}

By measuring the intrinsic speed of the bump under constant current injection, we observe that the change in the activity profile's width follows the relation in Eq.~\ref{eq:change-std} (Fig.~\ref{fig:UPI-results}A). Furthermore, the resulting intrinsic speed is in general well-predicted by Eq.~\ref{eq:new-speed}, elucidating the mechanism by which an external current can modulate the velocity of the attractor (Fig.~\ref{fig:UPI-results}B). A noticeable discrepancy emerges at higher positive values of $I_\text{speed}$, consistent with the deviation observed in Fig.~\ref{fig:beta-constants}A. This shared mismatch is driven by the inherent non-linearity of the activation function and the subsequent breakdown of the approximation: for sufficiently large speed currents, the firing rate function can no longer be linearized as formulated in Eq.~\ref{eq:linearized_req}.

Finally, Fig.~\ref{fig:UPI-results}C shows that in a dynamic scenario, where the input current varies with time to reflect the agent’s velocity, the system accurately tracks the true speed. This confirms that the trained recurrent layer functions as an effective unidirectional path integrator.

\section{\label{sec:discussion}Discussion}

In this work, we provided a unified account of how structured connectivity emerges from local plasticity and experience to produce representations of the input manifold capable of anticipation and path integration. We showed that Gaussian profiles in both feedforward and recurrent projections arise through self-organization within a competitive network. This competition is mediated by the dynamic interplay between global inhibition, firing-rate adaptation, and Hebbian learning.

In the absence of recurrent excitatory collaterals, the mathematical tractability of the feedforward framework permits the derivation of closed-form fixed-point solutions for the learned connectivity profiles and facilitates a rigorous stability analysis. However, when plastic recurrent excitatory connectivity is introduced, the analytical treatment becomes non-trivial. To address this, we employed extensive numerical simulations to characterize the stability conditions of the full architecture.

Specifically, we observed that the transition from a disordered state to a fully self-organized attractor requires a sufficiently strong tutor current. While previous models have demonstrated the emergence of attractor connectivity through learning \citep{widloski2014model}, our simplified framework offers deeper insights into the underlying stability regimes and the necessary conditions for robust self-organization in such systems.

A central finding of our study is that prospective coding can emerge not only within recurrent architectures—as extensively documented in previous literature \citep{mi2014spike,yu2019spiking,li2024dynamics}—but also through purely feedforward mechanisms. This agrees with the modeling work of \citet{treves2004computational}, who showed that firing-rate adaptation produces anticipation in networks with recurrent collaterals and in a downstream stage that lacks them, largely independently of the architecture. Our framework adds an analytical account of how the connectivity supporting this shift self-organizes from local plasticity, with closed-form solutions and stability conditions, and of how, once recurrent interactions are included, the same network supports path integration. Our finding that the predictive shift scales with adaptation strength matches the view that adaptation is the key ingredient, providing a candidate mechanism for the anticipatory firing observed in the superficial layers of the entorhinal cortex \citep{kropff2015speed,ouchi2024predictive}.

Furthermore, our results indicate that this effect can be amplified within hierarchical architectures, which could contribute to the layer-specific organization of prospective coding in the entorhinal cortex, where anticipatory signals are predominantly observed in superficial layers. Importantly, the entorhinal–hippocampal circuit should not be viewed as a simple feedforward stack, but rather as a distributed loop in which activity is transformed across multiple stages of processing and feedback \citep{basu2015corticohippocampal}. In this context, the anticipatory dynamics generated by our model can arise independently of recurrent excitatory attractor connections. This is compatible with the relative scarcity of such connectivity in superficial entorhinal layers, where interneuron-mediated inhibition plays a prominent role \citep{dhillon2000laminar,couey2013recurrent,pastoll2013feedback}. We stress, however, that this correspondence is suggestive rather than conclusive, and should not be extrapolated to other regions exhibiting anticipatory coding, where connectivity may differ.

Ultimately, for tasks like prospective coding, the functional differences between feedforward and recurrent architectures may be relatively minor within the class of models considered here, as the crucial computational ingredient is the intrinsic firing-rate adaptation \citep{treves2004computational}. Our findings, together with earlier work, therefore do not prescribe a strictly necessary anatomical connectivity structure: adaptation can drive anticipation across varied network topologies. Path integration, in contrast, requires accumulating velocity information over time, for which recurrent connectivity remains a viable mechanism, whereas feedforward projections may help propagate and shape predictive states across the entorhinal--hippocampal network, supporting anticipation regardless of which layer implements the integration.

Finally, we extended the model’s complexity by introducing a variable-speed signal. This velocity information was encoded via an additive global input current delivered to the network. Importantly, relying on an additive rather than multiplicative current makes this mechanism biologically plausible, while keeping our framework compatible with other established path integration models \citep{burak2009accurate,widloski2014model}. We showed that, upon the removal of the tutor location signal, the system is capable of path integration by modulating its intrinsic translation speed coherently with the speed signal. Our analysis revealed that this global current alters the steady-state solutions of the system by inducing a baseline shift in the Gaussian activity profiles. Due to the quadratic activation function, this baseline shift manifests as a modulation of the bump’s width proportional to the input current. This mechanism provides a robust biophysical account of speed modulation, suggesting how such architectures can maintain accurate spatial representations through path integration.

In summary, our work showed that path integration and prospective coding may not be two independent features of a neural system, but rather two features of the same self-organizing network. Our results are, however, restricted to an idealized setting in which each input unit has a single place field and field centers are homogeneously distributed along a one-dimensional manifold. Thus, the correlations among inputs reflect a clean one-dimensional structure. Starting from random initial projections, the connectivity is learned from this structure, but the structure itself, together with the single-field tuning of the inputs, is assumed. Empirical recordings show that spatially tuned neurons typically exhibit multiple, irregularly distributed firing fields in large environments \citep{park2011ensemble,eliav2021multiscale}, which is consistent with the non-topographical organization of their afferent inputs \citep{mainali2025universal}. Because this place-field disorder has important implications for both path integration and anticipatory coding \citep{schonsberg2024continuous}, evaluating how these computations emerge under more biologically realistic conditions is a natural next step. The analytical tractability of the present model provides a foundation for such extensions, as well as for extensions to two-dimensional manifolds. These could eventually contribute to a more complete account of how grid-cell representations are formed and used for spatial cognition.

\begin{acknowledgments}
This work was supported by the National Scientific and Technical Research Council of Argentina (CONICET). The authors extend their gratitude to the Leloir Institute for hosting this research and providing essential institutional support.
\end{acknowledgments}

\appendix

\section{\label{app:MathematicalBasis}Hermite Polynomials and the Projection Method}

\subsection{\label{appsub:Polynomial}The Classical Hermite Expansion}
Stable activity bumps emerge in networks with translationally invariant recurrent connectivity \cite{fung2008dynamics,fung2010moving}. While the cited framework accommodates general connectivity matrices, it is particularly tractable for Gaussian kernels. In a regime with no adaptation dynamics, perturbative analysis of the system leads to the following eigenvalue problem:

\begin{equation}
    \tau \frac{\partial }{\partial t}\delta U(x,z) = -\delta U(x,z) + \int dx^{\prime}\, F(x,x^{\prime}|z) \delta U(x^{\prime},z),\label{original_perturbative_solution}
\end{equation}

\noindent where the kernel of the perturbative operator $F(x,x^{\prime}|z)$ is defined as:

\begin{equation}
    \begin{split}
    F(x,x^{\prime}|z) =& \frac{2\rho_c U^{eq}(x^{\prime},z)}{B}\Bigg[W^{eq}(x,x^{\prime})\\
     &-  k\rho_c\int_{x^{\prime\prime}}dx^{\prime\prime} W^{eq}(x,x^{\prime\prime})r^{eq}(x^{\prime\prime},z)\Bigg].
    \end{split}\label{eq:perturbativeFung}
\end{equation}

\noindent Here, the normalization factor $B$ is given by $B = 1+k\rho_c\int dx [U^{eq}(x,z)]_+^2$, which simplifies to $1+k\rho_c A_u^2/(\sqrt{\pi}2\sigma_u)$ for Gaussian profiles $U^{eq}(x,z) = A_u \mathcal N(x;\,\,z,\sigma_u)$.

A key strength of this formalism is that $F(x,x^{\prime}|z)$ can be diagonalized using an orthonormal basis derived from Hermite polynomials. Consequently, any perturbation or shift in the Gaussian bump can be decomposed into a weighted sum of these orthogonal modes (Eq.~\ref{eq:decomposition}). The basis or right eigenfunctions $v_n(x|z)$ are defined as:

\begin{equation}
    v_n(x,\sigma_u) = \sqrt\frac{\pi^{1/2}2\sigma}{n!2^n}\mathcal N(x;\,\,z,\sigma_u) H_n\left(\frac{x-z}{\sigma_u}\right),
\end{equation}

\noindent where $H_n(x)$ denotes the $n$-th order physicist's Hermite polynomial, typically generated via Rodrigues' formula:

\begin{equation}
    H_n(x) = (-1)^n e^{x^2} \frac{d^n}{dx^n} e^{-x^2}.
\end{equation}

Under this decomposition, the first two excitation modes—representing the change in amplitude (zeroth-order) and the spatial shift (first-order) of the bump—are given by:

\begin{align}
    v_0(x|z) &= \sqrt{\pi^{1/2}2\sigma_u} \mathcal N (x;\,\, z, \sigma_u) \\
    v_1(x|z) &= 2\sqrt{\pi^{1/2}\sigma_u}\mathcal N (x;\,\, z, \sigma_u)\frac{x-z}{\sigma_u}.
\end{align}

\subsection{\label{appsub:ProjectionMethod}The Projection Method}

Since the steady-state bump profile can be expressed in terms of an orthonormal basis, a powerful method for analyzing the contributions of different modes is to propose a physically motivated ansatz. Typically, this takes the form of a Gaussian profile, $U^{eq}(x,z)$, potentially incorporating deformations corresponding to higher-order modes:

\begin{equation}
    U(x,z) = U^{eq}(x,z) + \sum_{n} c_n v_n(x|z). \label{eq:decomposition}
\end{equation}

The projection method consits of substituting the Gaussian ansatz into the governing dynamical equations, and projecting them onto the dominant spatial modes (typically the first two Hermite polynomials). Operationally, this involves multiplying both sides of the dynamical equations by a corresponding basis function $v_n(x|z)$ and integrating over the spatial domain $x$. This procedure significantly reduces the dimensionality of the system, yielding a tractable description governed by a small set of parameters.

Prior to deriving the projected equations for the various scenarios explored in this work, we establish several useful integral identities:

\begin{align}
    \int dx^{\prime}\, v_0(x^{\prime}|z) v_0(x^{\prime}|z)  &= 1, \label{eq:INT0} \\
    \int dx^{\prime}\, v_0(x^{\prime}|z) v_1(x^{\prime}|z) &= 0, \label{eq:INT1} \\
    \int dx^{\prime}\, v_1(x^{\prime}|z) v_1(x^{\prime}|z) &= 1, \label{eq:INT2} \\
    \int dx^{\prime}\, v_0(x^{\prime}|z) v_0(x^{\prime}|z-d) &= \exp\left({-\frac{d^2}{4\sigma_u^2}}\right),\label{eq:INT3} \\
    \int dx^{\prime}\, v_0(x^{\prime}|z) v_1(x^{\prime}|z-d) &= \frac{d}{\sqrt{2}\sigma_u}\exp\left({-\frac{d^2}{4\sigma_u^2}}\right) \label{eq:INT4}, \\
    \int dx^{\prime}\, v_1(x^{\prime}|z) v_1(x^{\prime}|z-d) &=  \left(1 - \frac{d^2}{2\sigma_u^2}\right)\exp\left({-\frac{d^2}{4\sigma_u^2}}\right). \label{eq:INT6}
\end{align}

\noindent These identities facilitate the application of the projection method to our two primary cases of study: 

\subsubsection{\label{app:EmergentAsym}Case I: $I(x,z) = \int \, dx^{\prime} J^{eq}(x,x^{\prime})R(x^{\prime},z)$}

We apply the projection method to the dynamical system defined in Eqs.~\ref{eq:U-dynamics}-\ref{eq:V-dynamics} using the following ansatz:

\begin{equation}
    \begin{aligned}
        U^{eq}_\gamma (x,z) &= \tilde{A}_u \left[v_0(x|z) + \tilde{\gamma} v_1(x|z)\right]
    \end{aligned}
\end{equation}
\begin{equation}
    \begin{aligned}
        V^{eq}_\gamma (x,z) &= \tilde{A}_ve^{\frac{y^2}{2}} v_0(x|z-\sqrt{2}\sigma_u y).
    \end{aligned}
\end{equation}

Note that this formulation is mathematically equivalent to the ansatz proposed in Eq.~\ref{eq:U-ansatz-1} via the following change of variables, introduced here for notational simplicity:

\begin{equation}
    \begin{aligned}
        u = \frac{\tau v}{\sqrt{2}\sigma_u}, \quad \Gamma = \frac{\tau_v}{\tau},& \quad y = \frac{d}{\sqrt{2}\sigma_u}, \quad \tilde{\gamma} = \gamma / \sqrt{2}, \\
        \tilde{A}_u = \frac{A_u}{\sqrt{\pi^{1/2}2\sigma_u}}, \quad \tilde{A}_v &= \frac{A_v}{\sqrt{\pi^{1/2}2\sigma_u}} \exp\left({-\frac{d^2}{4\sigma_u^2}}\right). 
    \end{aligned}
\end{equation}

This transformation is used in the subsequent derivations of the Appendix without altering the underlying results. The objective is to determine the state variables $\tilde{\gamma}$, $y$, $\tilde{A}_u$, and $\tilde{A}_v$. By projecting the system onto the first two eigenmodes and utilizing the identities in Eqs.~\ref{eq:INT0}--\ref{eq:INT6}, we obtain:

\begin{align}
    -\tilde{\gamma} u \tilde{A}_u &= -\tilde{A}_u - \tilde{A}_v + \rho_{in}A_JA_R \label{eq:RES1} \\
    u \tilde{A}_u &= -\tilde{\gamma} \tilde{A}_u + y \tilde{A}_v \label{eq:RES2} \\
    \Gamma u y \tilde{A}_v &= -\tilde{A}_v + m \tilde{A}_u \label{eq:RES3} \\
    \Gamma u(1-y^2)\tilde{A}_v &= y \tilde{A}_v + \gamma \tilde A_u.  \label{eq:RES4}
\end{align}

Solving this closed system yields the following expressions for the amplitudes:

\begin{equation}
    \begin{aligned}
        \tilde{A}_u &= \frac{\rho_{in}y A_R A_J}{y(1-\tilde \gamma u) +u+ \tilde{\gamma}} \\
        \tilde{A}_v &= \frac{\rho_{in}(u+\tilde \gamma) A_R A_J}{y(1-\tilde \gamma u) +u+ \tilde{\gamma}}.
    \end{aligned}
\end{equation}

The parameters $\tilde{\gamma}=\gamma/\sqrt2$ and $y$ are further detailed in Eq.~\ref{eq:gamma-y-expressions}.

\subsubsection{\label{subsubapp:ProjMethUPI}Case II: $I(x,z) = \int \, dx^{\prime} W^{eq}(x,x^{\prime})r(x^{\prime},z)  + \alpha(v-v_\text{int}(m))$}

For this case, we write the ansatz introduced in Eq.~\ref{eq:U-ansatz-1} as follows

\begin{align}
        U^{eq}_\epsilon (x,z) &= \tilde{A}_u v_0(x|z) + \epsilon_u I_\text{speed},\\
        V^{eq}_\epsilon (x,z) &= \tilde A_v e^{\frac{y^2}{2}}v_0(x|z-\sqrt{2}\sigma_u y) + \epsilon_v I_\text{speed}.
\end{align}

By assuming $\epsilon_u I_\text{speed}\ll \frac{A_u}{\sigma_u\sqrt{2\pi}}$ and projecting the system onto the first two modes one gets the following set of equations:

\begin{align}
    0 =& -\tilde A_u - \tilde A_v \nonumber\\
    &+ \frac{\rho_c A_W\tilde A_u^2}{\sqrt{2\sigma_u\sqrt{\pi}}B}\left(1+4\sqrt{\frac{2\sigma_u\sqrt{\pi}}{5}}\frac{\epsilon_u I_\text{speed}}{A_u}\right)\label{eq:RES21} \\
    u\tilde A_u =& y\tilde A_v \label{eq:RES22}\\
    \Gamma u y \tilde{A}_v =& -\tilde{A}_v + m \tilde{A}_u \label{eq:RES23} \\
    \Gamma u(1-y^2)\tilde{A}_v =& y \tilde{A}_v.  \label{eq:RES24}
\end{align}

By rearranging Eqs. \ref{eq:RES22}-\ref{eq:RES24} one gets

\begin{align}
    u &= \sqrt{\frac{m}{\Gamma}}y \label{eq:adim-vel}\\
    y &= \sqrt{1-\sqrt\frac{1}{\Gamma m}} \label{eq:adim-lag}
\end{align}

\begin{figure}[!ht]
    \centering
    \includegraphics[width=1.\linewidth]{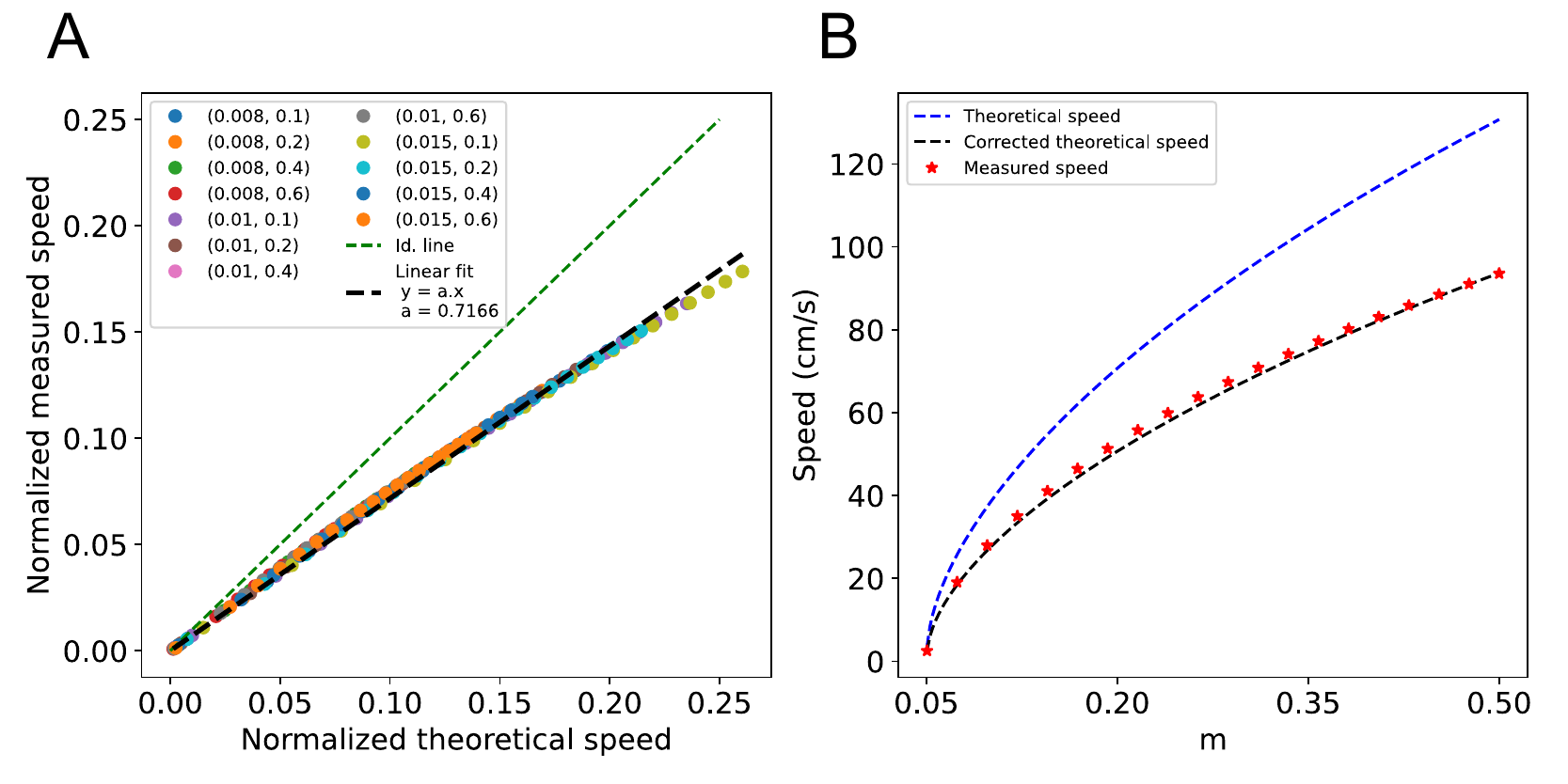}
    \caption{\textbf{Velocity formula comparison}. 
    \textbf A, Measured versus theoretical speeds are compared across different combinations of time constants ($\tau$ and $\tau_v$, in seconds) indicated in the legend. Both speeds are normalized by their respective $\sigma_u/\tau$ factors. Each data point represents a distinct $m$ value ranging from $1.01 \cdot \tau/\tau_v$ to 0.25. The green dashed line indicates the identity line, highlighting the discrepancy between theoretical and measured values. The black dashed line represents the linear fit, which yields a slope of $0.7166\pm0.0027$ (slope $\pm$ SE). Other simulation parameters: $N_c=1000$, $L=100$ cm, $\sigma_W=5\,\mathrm{cm}$, $k=1$ $\text{cm}^4\text{s}^2/\text{syn}$, and $A_W=1$ syn.
    \textbf B, Speed as a function of adaptation strength. Red stars denote simulation results obtained with the same parameters as panel A and specific time constants $\tau=15\,\mathrm{ms}$ and $\tau_v=600\,\mathrm{ms}$. The blue dashed curve shows the theoretical prediction from the projection method, while the black dashed curve corresponds to the empirically corrected expression, obtained by rescaling Eq.~\ref{eq:adim-vel} by a the estimated slope from the linear fit.}
    \label{fig:THEORETICALMISMATCH}
\end{figure}

It is straightforward to verify that Eqs. \ref{eq:adim-vel} and \ref{eq:adim-lag} recover the same expressions for the speed $v$ and lag $d$ as those derived in \cite{mi2014spike}. However, our simulation results reveal a systematic deviation from the theoretical speed given by Eq.~\ref{eq:theovel}. To quantify this discrepancy, we performed a linear regression of the simulated speeds against the theoretical predictions across various combinations of the parameters $\tau$ and $\tau_v$, with both speeds normalized by their corresponding ratio $\sigma_u/\tau$ (see Fig.~\ref{fig:THEORETICALMISMATCH}). The resulting fit yields a slope of $0.7166\pm0.0027$ (slope $\pm$ standard error), confirming a constant ratio mismatch where the original theory overestimates the actual speed.

To our knowledge, such a mismatch has not been previously reported in the literature. As resolving this issue lies outside the main scope of the present work, we adopt an empirically calibrated velocity law. Consequently, the theoretical results presented hereafter correspond to a corrected theoretical formula, obtained by multiplying the original expression by this $0.72$ factor. Nonetheless, we hypothesize that this underlying discrepancy comes from the Gaussian ansatz assumed for $V(x,z)$ in the projection method. Mathematically, $V(x,z)$ cannot be strictly Gaussian if $U(x,z)$ takes a Gaussian form (see Eq.~\ref{eq:V-dynamics}).

On the other hand, now amplitudes $\tilde A_u$ and $\tilde A_v$ become modulated by the input speed. By approximating $u$ by Eq.~\ref{eq:adim-vel} (neglecting the contributions induced by the changing standard deviation), and noting that $B = 1 + k\rho_c \frac{\tilde A_u^2}{2\sqrt{\pi}\sigma_u}$ we get

\begin{align}
    \tilde A_u &= \frac{-b + \sqrt{b^2-4ac}}{2a}, \\
    \tilde A_v &= \sqrt\frac{m}{\Gamma} \tilde A_u
\end{align}

\noindent with

\begin{align}
    a &= \left(1 + \sqrt\frac{m}{\Gamma}\right)\frac{k\rho_c}{2\sigma_u \sqrt \pi},\\
    b &= -\frac{\rho_c A_W }{\sqrt{2\sigma_u \sqrt \pi}},\\
    c &= \left(1 + \sqrt\frac{m}{\Gamma}\right) - \frac{4}{\sqrt 5} \rho_c A_W\epsilon_u I_\text{speed}.
\end{align}

\section{\label{app:FFWeightStability}Weight Stability and Hebbian Learning}

\subsection{The perturbative analysis}

In the main text, we have shown that translationally invariant input, coupled 
with Hebbian learning dynamics, can support both stable connectivity profiles and 
stable activity bumps. In this appendix, we develop a perturbative framework to 
analyze the stability of the learned feedforward synaptic weights. We assume 
interactions are mediated solely by global inhibition, neglecting recurrent 
excitatory collaterals. Our goal is to characterize the system's stability from 
a purely analytical perspective, complementing the simulation results and 
fixed-point solutions found earlier.

Assuming the stable weight matrix is well-approximated by a Gaussian profile, 
we consider small perturbations around this solution. Since $J^{eq}(x,x^{\prime})$ is Gaussian, 
the corresponding steady-state field $U^{eq}(x^{\prime},z)$ is also expected to be Gaussian. 
Consequently, a small perturbation $\delta J(x,x^{\prime})$ around the equilibrium induces a corresponding perturbation $\delta U(x,z)$ in the field. Linearizing the 
dynamics yields:

\begin{align}
    \frac{\partial}{\partial t} \delta J(x,x^{\prime}) =& \eta_J \delta r(x^{\prime},z) 
    \left[R(x,z)-\alpha_J J^{eq}(x,x^{\prime})^\beta\right]\nonumber\\
    &- \eta_J \beta \alpha_J r^{eq}(x^{\prime},z)J^{eq}(x,x^{\prime})^{\beta-1}\delta J(x,x^{\prime}), \\
    \tau\frac{\partial}{\partial t}\delta U(x^{\prime},z) =& - \delta U(x^{\prime},z) 
    - \delta V(x^{\prime},z) \nonumber\\
    &+ \rho_\text{in} \int dx\,R(x,z) \delta J(x,x^{\prime}) , \\
    \tau_v\frac{\partial}{\partial t}\delta V(x^{\prime},z) &= 
    -\delta V(x^{\prime},z) + m \delta U(x^{\prime},z).
\end{align}

First, we rewrite the perturbation in $U$ as:

\begin{align}
    \delta U(x^{\prime},z) =& \frac{1}{m+1}\left[-\tau \frac{\partial}{\partial t}
    \delta U(x^{\prime},z) \right. \nonumber\\
    &\left.+ \tau_v \frac{\partial}{\partial t}\delta V(x^{\prime},z) 
    + \rho_\text{in} \int R(x,z)\delta J(x,x^{\prime})\, dx\right].
    \label{eq:deltaU}
\end{align}

Second, we consider a periodic stimulus that repeats with period $T=L/\dot z$. 
Since Hebbian learning is much slower than the bump dynamics ($1/\eta_J \gg \tau_v$), 
we can average Eq.~\ref{eq:hebbian-learning} over this period, which is 
equivalent to averaging over all possible stimulus positions:

\begin{align}
    \frac{\partial }{\partial t}\delta J(x,x^{\prime}) \approx& \eta_J \left\langle 
    \left[R(x,z)-\alpha_J J^{eq}(x,x^{\prime})^\beta\right]\delta r(x^{\prime},z)\right\rangle_z \nonumber \\
    &- \eta_J\beta\alpha_J\langle r^{eq}(x^{\prime},z)\rangle_z J^{eq}(x,x^{\prime})^{\beta-1}\delta J(x,x^{\prime}).
    \label{eq:avgPerturbation}
\end{align}

We can express $\delta r(x,z)$ in terms of $\delta U(x,z)$ as:

\begin{align}
    \delta r(x^{\prime},z) =& \frac{2U^{eq}(x^{\prime},z)}{B}\delta U(x^{\prime},z)\nonumber\\ 
    &- \frac{2k\rho_\text{c} U^{eq}(x^{\prime},z)^2}{B^2}\int dx^{\prime\prime\prime}\,\,U^{eq}(x^{\prime\prime\prime},z)\delta U(x^{\prime\prime\prime},z).
    \label{eq:deltaR}
\end{align}

\noindent where $B=1+\frac{k\rho_c A_u^2}{2\sqrt\pi\sigma_u}$ takes the value corresponding to the Gaussian ansatz.

Substituting Eq.~\ref{eq:deltaR} into Eq.~\ref{eq:avgPerturbation}, rewriting 
$\delta U$ using Eq.~\ref{eq:deltaU}, and neglecting terms of order $\mathcal O(\eta_J\tau_v)$, 
we obtain:

\begin{widetext}
    \begin{equation}
        \begin{aligned}
            \frac{\partial }{\partial t}\delta J (x,x^{\prime}) &\approx 
            \frac{2\eta_J\rho_\text{in}}{B(m+1)}\int dx^{\prime\prime}\Bigg\langle 
            \left[R(x,z)-\alpha_J J^{eq}(x,x^{\prime})^\beta\right] \quad \times \Bigg[R(x^{\prime\prime},z) U^{eq}(x^{\prime},z)\delta J(x^{\prime\prime},x^{\prime}) \\
            &\quad - k\rho_c r^{eq}(x^{\prime},z)\int dx^{\prime\prime\prime} U^{eq}(x^{\prime\prime\prime},z) R(x^{\prime\prime},z)
            \delta J(x^{\prime\prime},x^{\prime\prime\prime})\Bigg]\Bigg\rangle_z \quad - \eta_J\beta\alpha_J\langle r^{eq}(x^{\prime},z)\rangle_z 
            J^{eq}(x,x^{\prime})^{\beta-1}\delta J(x,x^{\prime}).
        \end{aligned}
        \label{eq:combined_integro_diff}
    \end{equation}
\end{widetext}

Finally, we rewrite the differential equation by defining the kernels 
$T_1, T_2,$ and $T_3$:

\begin{widetext}
    \begin{align}
        \frac{\partial}{\partial t} \delta J(x,x^{\prime}) &=\underbrace{\frac{2\eta_J\rho_\text{in}}{B(m+1)}\int dx^{\prime\prime} \int dz 
        \left\{ [T_1(x,x^{\prime},x^{\prime\prime},z)-T_2(x,x^{\prime},x^{\prime\prime},z)]\delta J(x^{\prime\prime},x^{\prime}) 
        \right\}}_{\text{Term I}}\nonumber \\
        &\quad -\underbrace{\frac{k\rho_c}{2\sigma_u\sqrt{\pi}}\int dx^{\prime\prime}
        \int dx^{\prime\prime\prime}\int dz 
        \left\{ T_3(x^{\prime},x^{\prime\prime},x^{\prime\prime\prime},z)\delta J(x^{\prime\prime},x^{\prime\prime\prime}) \right\}}_{\text{Term II}} \quad -\underbrace{\eta_J\beta\alpha_J A_r J^{eq}(x,x^{\prime})^{\beta-1}
        \delta J(x,x^{\prime})}_{\text{Term III}},
    \end{align}
\end{widetext}

\noindent where:

\begin{widetext}
\begin{align}
    T_1(x,x^{\prime},x^{\prime\prime},z) =& A_R^2 A_u \mathcal N(x;\,\, z,\sigma_R)
    \mathcal N(x^{\prime\prime};\,\, z,\sigma_R)\mathcal N(x^{\prime};\,\, z,\sigma_u), \\
    T_2(x,x^{\prime},x^{\prime\prime},z) =& \alpha_J A_R A_u A_J^\beta\sqrt{\frac{(2\pi\sigma_J^2)^{1-\beta}}{\beta}} \mathcal N(x^{\prime\prime};\,\, z,\sigma_R) \mathcal N(x^{\prime};\,\, z,\sigma_u)
    \mathcal N\left(x;\,\, x^{\prime},\sigma_J/\sqrt{\beta}\right), \\
    T_3(x^{\prime},x^{\prime\prime},x^{\prime\prime\prime},z) =& A_r A_u A_R 
    \mathcal N\left(x^{\prime};\,\, z,\sigma_u/\sqrt{2}\right)
    \mathcal N(x^{\prime\prime};\,\, z,\sigma_R)
    \mathcal N(x^{\prime\prime\prime};\,\, z,\sigma_u).
\end{align}
\end{widetext}

In order to understand the stability of the system, we proceed to analyze 
the perturbative equations term by term.

\subsubsection{Term I}

We now analyze the structure of the kernel operators, focusing first on $T_1$.\\

\noindent \textbf{Operator $T_1$}\\

By completing the square in the exponent, we can express the operator as:

\begin{align}
    T_1(x,x^{\prime},x^{\prime\prime},z) =& \frac{A_R^2 A_u}{2\pi\sigma_R^2\sigma_u} 
    \sqrt{\frac{\sigma_u^2\sigma_R^2}{2\sigma_u^2+\sigma_R^2}} 
    \nonumber\\ &\times\exp\left[-\phi(x,x^{\prime},x^{\prime\prime})\right] 
    \mathcal{N}\left(z; \mu_z, \Sigma_z \right),
    \label{eq:T1_structure}
\end{align}

\noindent where the variance and mean of the Gaussian part are 
$\Sigma_z^2 = \frac{\sigma_u^2\sigma_R^2}{2\sigma_u^2+\sigma_R^2}$ 
and $\mu_z = \Sigma_z^2 G$, respectively. The exponential argument $\phi$ 
and the term $G$ are given by:

\begin{equation}
    \begin{aligned}
        \phi(x,x^{\prime},x^{\prime\prime}) =& \frac{x^2}{2\sigma_R^2} + \frac{{x^{\prime\prime}}^2}{2\sigma_R^2} 
        + \frac{{x^{\prime}}^2}{2\sigma_u^2} 
        \\ &- \frac{\sigma_u^2\sigma_R^2}{2\sigma_u^2+\sigma_R^2}
        \left(\frac{\sigma_u^2(x+x^{\prime\prime})+\sigma_R^2 x^{\prime}}{\sigma_R^2\sigma_u^2}
        \right)^2, \\
        G =& \frac{x}{\sigma_R^2} + \frac{x^{\prime}}{\sigma_u^2} 
        + \frac{x^{\prime\prime}}{\sigma_R^2}.
    \end{aligned}
    \label{eq:phi_B_def}
\end{equation}

Integrating Eq.~\ref{eq:T1_structure} over $z$ eliminates the normalized 
Gaussian factor $\mathcal{N}\left(z;\, \mu_z, \Sigma_z \right)$, reducing 
$T_1$ to the effective kernel $\mathcal T_1$:

\begin{align}
    \mathcal T_1(x,x^{\prime},x^{\prime\prime}) &\equiv \int dz \, T_1(x,x^{\prime},x^{\prime\prime},z) \\
    &= \frac{A_R^2 A_u}{2\pi\sigma_R^2\sigma_u} 
    \sqrt{\frac{\sigma_u^2\sigma_R^2}{2\sigma_u^2+\sigma_R^2}} 
    \exp\left[-\phi(x,x^{\prime},x^{\prime\prime})\right].
    \label{eq:T1_integrated}
\end{align}

We can rewrite the quadratic form $\phi$ explicitly as:

\begin{equation}
    \begin{aligned}
        \phi(x,x^{\prime},x^{\prime\prime}) =& \frac{\sigma_u^2+\sigma_R^2}
        {2\sigma_R^2(\sigma_R^2+2\sigma_u^2)}x^2 
        \\
        &+\frac{\sigma_u^2+\sigma_R^2}
        {2\sigma_R^2(\sigma_R^2+2\sigma_u^2)}{x^{\prime\prime}}^2 + \frac{1}{2\sigma_u^2+\sigma_R^2}{x^{\prime}}^2 \\
        &- \frac{1}{2\sigma_u^2+\sigma_R^2}
        \left(xx^{\prime} + x^{\prime\prime}x^{\prime} + \frac{\sigma_u^2}{\sigma_R^2}xx^{\prime\prime}\right).
    \end{aligned}
    \label{eq:phi_explicit}
\end{equation}

To exploit the translational invariance, we introduce the relative coordinates:

\begin{equation}
    s = x-x^{\prime}, \quad s^{\prime\prime}=x^{\prime\prime}-x^{\prime}.
    \label{eq:change_vars}
\end{equation}

Substituting these into Eq.~\ref{eq:phi_explicit}, $\phi$ becomes dependent 
solely on the relative distances $s$ and $s^{\prime\prime}$:

\begin{equation}
    \phi(s,s^{\prime\prime}) = \frac{1}{2\sigma_R^2(2\sigma_u^2+\sigma_R^2)}
    \left[(s^2+{s^{\prime\prime}}^2)(\sigma_u^2+\sigma_R^2) + 2ss^{\prime\prime}\sigma_u^2\right].
    \label{eq:phi_relative}
\end{equation}

Thus, the final form of the integrated Term I kernel is:

\begin{equation}
    \mathcal T_1(s,s^{\prime\prime}) = \frac{A_R^2 A_u e^{-\phi(s,s^{\prime\prime})}}
    {2\pi\sigma_R \sqrt{2\sigma_u^2+\sigma_R^2}}.
    \label{eq:T1_final}
\end{equation}

\noindent \textbf{Operator $T_2$}

For this second operator, we define the projection against $z$ in an analogous way:

\begin{equation}\begin{aligned}
\mathcal T_2 (x,x^{\prime},&x^{\prime\prime}) \equiv \int_z dz T_2(x,x^{\prime},x^{\prime\prime},z)\\
=&\alpha_J A_J^\beta C_\beta A_R A_u \mathcal N\left(x;\,\,x^{\prime},
\frac{\sigma_J}{\sqrt{\beta}}\right)\\
&\times\int_{z}dz\mathcal N(x^{\prime\prime};\,\, z,\sigma_R)\mathcal N(x^{\prime};\,\, z,\sigma_u)\\
=& A_R^2 A_u \mathcal N\left(x;\,\, x^{\prime},\frac{\sigma_J}{\sqrt{\beta}}\right)
\mathcal N\left(x^{\prime};\,\, x^{\prime\prime},\sqrt{\sigma_R^2 + \sigma_u^2}\right),
\end{aligned}\end{equation}

\noindent with $C_\beta \equiv \sqrt{\frac{(2\pi\sigma_J^2)^{1-\beta}}{\beta}}$ as defined in 
Eq.~\ref{eq:AJ-Cbeta}.

This kernel also has translation invariance with respect to $x^{\prime}$, 
so we can rewrite it as:

\begin{equation}
    \mathcal T_2(s,s^{\prime\prime}) =  A_R^2 A_u \mathcal N \left(s;\,\, 0,
    \frac{\sigma_J}{\sqrt{\beta}}\right)
    \mathcal N \left(s^{\prime\prime};\,\,0,\sqrt{\sigma_R^2 + \sigma_u^2}\right).
\end{equation}

\subsubsection{Term II (Global Inhibition)}

The second term (Term II) involves a double spatial integral and represents 
global interactions mediated by the inhibition strength $k$. Because this 
term couples all perturbations across the domain non-locally, an analytical 
treatment is not straightforward. Nevertheless, the overall minus sign 
suggests that global inhibition can act as an additional damping mechanism 
for certain classes of perturbations, although in general its effect need 
not be sign-definite for arbitrary sign-changing eigenmodes. In what follows, 
since Term II is directly proportional to $k$ and we focus on the 
weak-inhibition regime, we neglect this contribution at leading order. However, as shown in Fig.~\ref{fig:TheoreticalStabilityLambdas}, simulation results indicate that increasing the inhibition factor does not affect stability.

\subsection{The resulting generalized eigenvalue-eigenfunction problem}

By neglecting Term II, the perturbation dynamics are governed by Term I 
(Hebbian covariance operators) and Term III (local leak). It is convenient 
to write the linearized learning dynamics in the compact form

\begin{align}
    \tau_J(x,x^{\prime})&\,\frac{\partial}{\partial t}\,\delta J(x,x^{\prime},t)\nonumber
    \\ & = \int dx^{\prime\prime}\,K(x,x^{\prime},x^{\prime\prime})\,\delta J(x^{\prime\prime},x^{\prime},t) - \delta J(x,x^{\prime},t),
    \label{eq:simplified_dynamics}
\end{align}

\noindent where the effective learning time constant is
\begin{equation}
    \tau_J(x,x^{\prime}) \equiv \frac{J^{eq}(x,x^{\prime})^{1-\beta}}{\beta\eta_J\alpha_J A_r},
\end{equation}

\noindent and the linear kernel is
\begin{align}
    K(x,x^{\prime},x^{\prime\prime}) \equiv& \frac{2\rho_\text{in}}{B(m+1)}\,\frac{J^{eq}(x,x^{\prime})^{1-\beta}}
    {\beta\alpha_J A_r}\,\nonumber
    \\ &\times\Big[\mathcal{T}_1(x,x^{\prime},x^{\prime\prime}) - \mathcal{T}_2(x,x^{\prime},x^{\prime\prime})\Big].
\end{align}

Since $J^{eq}(x,x^{\prime}) = A_J \mathcal{N}(x; \,\,x^{\prime}, \sigma_J)$ depends only on $s = x - x^{\prime}$, both $\tau_J$ and $K$ are translation invariant with 
respect to $x^{\prime}$, i.e. $\tau_J(x,x^{\prime}) = \tau_J(s)$ and 
$K(x,x^{\prime},x^{\prime\prime}) = {K}(s, s^{\prime\prime})$. 
Eq.~\ref{eq:simplified_dynamics} can therefore be rewritten as

\begin{align}
    \tau_J(s)&\,\frac{\partial}{\partial t}\,\delta J(x,x^{\prime},t)\nonumber\\
    &= \int ds^{\prime\prime}\,\tilde{K}(s,s^{\prime\prime})\,\delta J(s^{\prime\prime}+x^{\prime},x^{\prime},t) - \delta J(x,x^{\prime},t).
\end{align}

Because neither $\tilde{K}(s,s^{\prime\prime})$ nor $\tau_J(s)$ depend on $x^{\prime}$, the 
spectral structure of the problem is identical for every value of $x^{\prime}$. 
Therefore, we may fix an arbitrary $x^{\prime}$ and define 
$w_{x^{\prime}}(s,t) \equiv \delta J(s+x^{\prime}, x^{\prime}, t)$, reducing the dynamics to

\begin{align}
    &\tau_J(s)\,\frac{\partial}{\partial t}\,w_{x^{\prime}}(s,t)
    = \hat{L}[w_{x^{\prime}}(s,t)], \nonumber
    \\ &\text{with} \quad 
    \hat{L}[w] \equiv \int ds^{\prime\prime}\,\tilde{K}(s,s^{\prime\prime})\,w(s^{\prime\prime},t) - w(s,t).
\end{align}

Since $\hat{L}$ is a linear operator and $\tau_J(s) > 0$ for all $s$, 
we seek solutions of the form

\begin{equation}
    w_{x^{\prime}}(s,t) = \sum_i c_i(x^{\prime})\, f_i(s)\, e^{\lambda_i t},
\end{equation}

\noindent where the coefficients $c_i(x^{\prime})$ are determined by the initial condition 
$w_{x^{\prime}}(s,0)$ and carry all the $x^{\prime}$-dependence of an arbitrary perturbation. 
We note that the kernel $\tilde{K}(s,s^{\prime\prime})$ is not symmetric under exchange 
of $s$ and $s^{\prime\prime}$ (since $J(s)^{1-\beta}$ depends on $s$ but not $s^{\prime\prime}$), 
so $\hat{L}$ is not self-adjoint. Accordingly, we assume that $\hat{L}$ is 
diagonalizable, i.e.\ that its eigenfunctions $\{f_i(s)\}$ form a complete 
basis; should this fail, generalized eigenfunctions with algebraic 
$t e^{\lambda_i t}$ time dependences would arise, but the stability 
criterion derived below would remain unchanged. Under this assumption, 
substituting the ansatz yields the generalized eigenvalue-eigenfunction 
problem for each mode:

\begin{equation}
    \hat{L}[f_i(s)] = \lambda_i\,\tau_J(s)\,f_i(s),
\end{equation}

The system is stable if and only if $\mathrm{Re}(\lambda_i) < 0$ for all 
$i$. Since the spectrum is independent of $x^{\prime}$, this criterion constitutes 
a global stability condition in the weak-inhibition regime.

\subsection{Discretization and Numerical Stability Analysis}

To assess stability numerically, we discretize the domain of $s$ and $s^{\prime\prime}$ 
over a uniform grid of $M$ points with spacing $\Delta s$. Under this 
discretization, the continuous integral operator is approximated by a 
Riemann sum, so the continuous operators become:
\begin{itemize}
    \item $\tilde{K}(s, s^{\prime\prime})$ $\rightarrow$ an $M\times M$ matrix 
    $\mathbf{K}$, with entries 
    $\mathbf{K}_{ij} = \tilde{K}(s_i, s_j)\,\Delta s$,
    \item $\tau_J(s)$ $\rightarrow$ a diagonal $M\times M$ matrix 
    $\boldsymbol{\tau}$, with entries $\boldsymbol{\tau}_{ii} = \tau_J(s_i)$,
    \item $\hat{L}$ $\rightarrow$ the matrix 
    $\mathbf{L} \equiv \mathbf{K} - \mathbf{Id}$,
\end{itemize}
where $\mathbf{Id}$ is the $M\times M$ identity matrix. Note that the 
quadrature weight $\Delta s$ must be absorbed into $\mathbf{K}$ to ensure 
dimensional consistency of the discrete approximation. The generalized 
eigenvalue problem then becomes the matrix eigenvalue problem

\begin{equation}
    \boldsymbol{\tau}^{-1}\,\mathbf{L}\,\mathbf{f}_i = -\lambda_i\,\mathbf{f}_i,
    \label{eq:discrete_eigen}
\end{equation}

\noindent where $\mathbf{f}_i$ is the $i$-th eigenvector and $-\lambda_i$ is the 
corresponding eigenvalue of $\boldsymbol{\tau}^{-1}\mathbf{L}$. Since 
$\boldsymbol{\tau}$ is diagonal with strictly positive entries, its inverse 
is well defined. The stability criterion $\Re(\lambda_i) > 0$ for 
all $i$ translates directly into the requirement that all eigenvalues 
of $\boldsymbol{\tau}^{-1}\mathbf{L}$ have strictly negative real part. 
Stability analysis therefore reduces to computing the eigenspectrum of the 
matrix $\boldsymbol{\tau}^{-1}\mathbf{L}$, which is the problem we solve numerically.

\bibliography{bibliography}

\end{document}